\documentclass{./ieeeaccess}
\usepackage[english]{babel}
\usepackage[utf8]{inputenc}

\usepackage{cite}
\usepackage{amsmath,amssymb,amsfonts}
\usepackage{graphicx}
\usepackage{textcomp}
\def\BibTeX{{\rm B\kern-.05em{\sc i\kern-.025em b}\kern-.08em
    T\kern-.1667em\lower.7ex\hbox{E}\kern-.125emX}}

\usepackage{booktabs,enumerate,multirow} 

\usepackage{url}

\usepackage{algpseudocode}

\usepackage[section]{placeins}

\newcounter{MYtempeqncnt}

\newtheorem{proposition}{Proposition}
\newtheorem{lemma}{Lemma}

\newcommand{\Dparcial}[2]{\frac{\partial #1}{\partial #2}}

\newcommand{\cardB}{B}
\newcommand{\cardS}{S}
\newcommand{\ncap}[1]{\gamma^{(#1)}}
\newcommand{\suscRatio}[1]{\sigma^{(#1)}}
\newcommand{\penratio}[1]{\sigma^{(#1)}}

\let\vec\mathbf

\usepackage{calc}

\begin{document}

\history{\copyright 2021 IEEE.  Personal use of this material is permitted.  Permission from IEEE must be obtained for all other uses, in any current or future media, including reprinting/republishing this material for advertising or promotional purposes, creating new collective works, for resale or redistribution to servers or lists, or reuse of any copyrighted component of this work in other works.}
\doi{10.1109/ACCESS.2021.3078562}

\title{Competition between service providers with strategic resource allocation: application to network slicing}

\author{\uppercase{Luis Guijarro}\authorrefmark{1},\uppercase{Jose~R. Vidal}\authorrefmark{2}, and \uppercase{Vicent Pla}\authorrefmark{3}}
\address[1]{Universitat Polit\`ecnica de Val\`encia, Spain (e-mail:lguijar@dcom.upv.es)}
\address[2]{Universitat Polit\`ecnica de Val\`encia, Spain (e-mail:jrvidal@dcom.upv.es)}
\address[3]{Universitat Polit\`ecnica de Val\`encia, Spain (e-mail:vpla@upv.es)}

\corresp{Corresponding author: Luis Guijarro (e-mail:lguijar@dcom.upv.es)}

\tfootnote{This work has been supported by the Spanish Ministry of Science and Universities (MCIU/AEI) and the European Union (FEDER/UE) through Grants PGC2018-094151-B-I00 and RED2018-102585-T.}

\begin{abstract}
We propose and analyze a business model for a set of operators that use the same physical network.   
Each operator is entitled to a share of a network operated by an Infrastructure Provider (InP) and uses network slicing mechanisms to request network resources as needed for service provision. The network operators become Network Slice Tenants (NSTs).
The InP performs the resource allocation based on a vector of weights chosen selfishly by each NST. The weights distribute the NST's share of resources between its subscribers in each cell.
We model this relationship as a game propose a solution for the Nash equilibrium in which each NST chooses weights equal to the product of its share by the ratio between the total number of subscribers in the cell and the total number of subscribers in the network.
We characterize the proposed solution in terms of subscription ratios and fractions of subscribers, for different cell capacities and user sensitivities.
The proposed solution provides the exact values for the Nash equilibrium if the cells are homogeneous in terms of normalized capacity, which is a measure of the total amount of resources available in the cell. 
Otherwise, if the cells are heterogeneous, it provides an accurate approximation. We quantify the deviation from the equilibrium and conclude that it is highly accurate.

\end{abstract}


\maketitle
\section{Introduction}

\PARstart{T}{he} current mobile network architecture utilizes a relatively monolithic access and transport framework to accommodate a variety of services such as mobile traffic for smartphones, OTT content, feature phones, data cards, and embedded M2M devices. It is anticipated that this architecture will not be flexible and scalable enough to support the coming services, which demand very diverse use cases and sometimes extreme requirements---in terms of performance,  scalability and availability. Furthermore, the introduction of new network services should be made more efficiently~\cite{ngmn2016}. 

In the above scenario, network slicing is gaining an increasing importance as an effective way to introduce flexibility in the management of network resources. 
A network slice is a collection of network resources and functions, selected in order to satisfy the requirements  of the service(s) to be provided by the slice in terms of both QoS and functionality. 
Moreover, resource allocation to slices must be done in a way that ensures a certain degree of isolation between slices
so that traffic variations in one slice do not negatively impact the performance of other slices.
An enabling aspect of network slicing is virtualization.  Virtualization of network resources allows operators to share the same physical resource in a flexible and dynamic manner to exploit the available resources more efficiently, thus allowing to support a higher traffic load with the same amount of resources~\cite{jiang2016}.


Within the above context, we envision a scenario where a set of network operators use network slicing mechanisms to request network resources as needed for service provision. The InP is responsible for the network operation and maintenance, while the network operators become Network Slice Tenants (NSTs)%
\footnote{We can also envision other emergent players becoming NSTs, e.g., OTT service providers. Even Vertical Industry players may take this role when needing connectivity services, e.g., automotive verticals providing entertainment onboard, assisted driving and passenger safety; or traffic agencies needing traffic road state monitoring~\cite{samdanis2016,ngmn2015,elayoubi2019}.}.
The NSTs are entitled to a share of the network resources. This entitlement may result from diverse scenarios, e.g., the operators owned the networks and decided to pool the networks and to outsource their operation to an InP.%
 
We propose a business model where
the NSTs provide service to end users. This service may be characterized by a series of performance constraints (e.g., transmission rate and delay) and each NST gets revenues from its subscribers.
In order to support the service, the NSTs request dynamically access and core network resources from an InP. The InP works as a supporting unit to the NSTs.
As a motivation example, we describe the scenario and the model in terms of mobile service operators and radio access networks.
In this example, \emph{resources} would refer to the radio spectrum in each cell of the network and the equipment necessary to transmit over it.


Our work is related to several other works on resource allocation between operators within the context of network sharing. 
There is a subset of these works, e.g.,  \cite{oladejo2017}, in which the operators take no decision at all, and it is the infrastructure owner who allocates resources among the operators in order to maximize some sort of network utility.
In another subset, e.g.,  \cite{caballero2019}, the operators do take decisions that influence the resource allocation, and the decisions are of a strategic nature, i.e., each decision influences and it is influenced by each other decision. However, these works fail to model the concurrent strategic interaction that is also present when the operators compete against each other for providing service to the users. 
We aim to address this limitation by explicitly modeling the simultaneous strategic interaction between operators in both the resource allocation and the service provision ``markets''.
The above two works are discussed in Section~\ref{sec:related}.

Our work analyzes how independent NSTs compete against each other following the business model described above. 
Our contributions are the following ones:
\begin{enumerate}
\item We show how the strategic interaction between the NSTs both in the provision of the service and in the slicing of the network can be modeled as a game, where the strategy is a weighted distribution of the NST's share of the resources between the cells.

\item We propose a solution for the Nash equilibrium in which the NST chooses weights equal to the product of its share by the ratio between the total number of subscribers in the cell and the total number of subscribers in the network.
This equilibrium results in each NST using a fraction of the resources in each cell equal to its share,
not as the result of a centralized decision made by the InP, but as the result of a game where each NST acts selfishly.

\item We characterize the proposed solution for different cell capacities and user sensitivities.

\item We prove that the proposed solution provides the exact values at the equilibrium if the cells are homogeneous in terms of normalized capacity, which is a measure of the available resources in the cell normalized to the service price, the number of users and the no-subscription option valuation. 
Otherwise, if the cells are heterogeneous, it provides an accurate approximation of the equilibrium. We quantify the deviation from the equilibrium and conclude that it is highly accurate.

\end{enumerate}


The paper is structured as follows.  In Section~\ref{sec:related}, the related work is presented and discussed. In Section~\ref{sec:model}, the model for the NSTs, the users, and the InP is described. In Section~\ref{sec:analysis}, a strategic game is formulated for the interaction between the NSTs, a solution for the Nash equilibrium is proposed and its exactness is discussed. In Section~\ref{sec:setup}, we describe the experimental setup. In Section~\ref{sec:results}, we characterize the solution and quantify the exactness of the proposed solution. And finally, Section~\ref{sec:conclusion} draws the conclusions.

\section{Related work}\label{sec:related}

Our work focuses on the strategic interaction that takes place between tenants in a network slicing scenario. In our work, this interaction between the NSTs is double-sided: there is one interaction due to the competition for the service provision to the users; and another one due to the procurement of resources from the InP.

As regards the service competition, this work draws on previous works by the authors on the provision of services based on different kinds of resources. For example, 
 in~\cite{guijarro2017jsac}, the utility that users derive depends on the sensing rate that each service provider is able to acquire and use for composing the service it offers.  And in~\cite{guijarro2019fgcs}, the net data rate that each service provider is able to acquire for building a data-based service determines how attractive the service is for the users.

As regards the resource procurement, we borrow the ideas from the Fisher market model, which provides us with the rationale for the resource allocation rule, as described in Section~\ref{sec:systemmodel}. The model basically comprises a set of buyers aiming to purchase multiple goods in a way that maximizes their utility subject to budget constraints.
This model has also been borrowed by recent proposals on resource allocation mechanisms. In~\cite{feldman2009}, the allocation of computational resources to users is analyzed and the budget may have a monetary interpretation. And in~\cite{caballero2019}, radio network resources are allocated to network slices, and the budget derives from an initial lease of resources from the NSTs to a common network pool operated by the InP. 

Apart from the proposals based on the Fisher market, there are other proposals for allocating resources to service providers (e.g.,  MVNOs,  tenants,  SaaS providers,  \ldots).
A set of proposals comes from the Kelly mechanism, which is a mechanism for allocating resources in a network originally proposed by F.~Kelly~\cite{kelly1997, kelly1998}, within the context of rate allocation and control in communication networks. It resembles the Fisher market in that the users also aim to maximize the utility when buying the resources. Here, however, there is no per-user budget constraint, but only an aggregate resource constraint. It has been shown that an efficient resource allocation can be achieved using a pricing mechanism known as Kelly mechanism, under the assumption that the users are price-takers.
The Kelly mechanism has been borrowed by resource allocation mechanisms for cloud computing~\cite{ma2010} and for network slicing~\cite{tun2019}. 
Another set of proposals incorporate auctions. Reference~\cite{tang2012} proposes first-price auctions and VCG auctions for allocating a resource to service providers, while~\cite{zhu2016} proposes a combinatorial auction that borrows elements from the VCG auction to allocate resources bundles to MVNOs. 
In both sets of proposals, the service providers interact strategically, as in our work. However, the resources are owned by the InP, who collects revenues from the service providers, either through a pricing mechanism (\cite{ma2010,tun2019}) or through an auction (\cite{tang2012,zhu2016}). These scenarios differ from our work, from~\cite{feldman2009} and from~\cite{caballero2019}. In these three works, the resources are not owned by the InP, but by the tenants themselves, who pool the resources and entrust their operation to the InP. There is no revenue to collect by the InP, but only to respect each tenant's contribution share.
Finally, there are some proposals for resource allocation where the InP maximizes the overall network rate, under the quality constraints imposed by the users and the MVNOs, like in~\cite{oladejo2017,mahindra2013}. In these proposals, however, the MVNOs do not take any decision, which is an unrealistic scenario viewed from our approach.

All in all, our work resembles~\cite{caballero2019} as regards the resource procurement is concerned. First, the tenants own the infrastructure and entrust their operation to the InP. And second, the allocation is of a hierarchical nature, where the tenants are involved in the allocation---as opposed to a centralized scheme, where the InP decides and executes the complete allocation to all tenants. Within this framework, the Fisher market is an appropriate model for the allocation of resources between the tenants.
However, our work differs importantly from~\cite{caballero2019} in that we model the tenants and the users as different agents with their particular incentives, which are the profits and the user utility, respectively. In~\cite{caballero2019}, each tenant operates as a proxy of its subscribers; this may fail to properly model the tenant incentives and the corresponding  business model. This difference has also an important implication in the user behavior modeling: while in our work the number of subscribers for each tenant depends on the tenant allocation decision, in~\cite{caballero2019} the number of subscribers is independent from the decision, since the number of subscribers is fixed \emph{a priori} as a parameter.

The adoption of game theory by works on resource allocation in the context of network slicing is currently popular. Two references that exemplify this trend are~\cite{dawaliby2019} and~\cite{doro2018}. These two works use game theory as a means to distribute a (resource allocation) optimization problem among the relevant agents: the IoT gateways (GWs) in~\cite{dawaliby2019} and the MVNOs in~\cite{doro2018}, since the centralized version of the resource allocation problem is not realistic and/or computationally tractable. Nevertheless, there are some important differences between these two references and our manuscript.
Firstly, the agents (GWs and MVNOs, respectively) do not compete for the users (IoT devices and mobile users, MUs, respectively), but they are given instead as model parameters (fixed and random, respectively), whereas in our work, the agents (NSTs) compete strategically for the users.
And secondly, the formulated games (cooperative coalitional games and non-cooperative congestion games, respectively) are used as a means for distributing an algorithm, while in our work, the game is the model, and we do not prescribe any algorithm.

This work takes our analysis and results in~\cite{guijarro2018electronics} as a starting point, and it goes far beyond them, since they are preliminary and not analytical.  Indeed, the model described in~\cite{guijarro2018electronics} does not contemplate the possibility of a no-subscription option for the users, and does not provide an analytical expression for the Nash equilibrium (and, consequently, does not provide any solution characterization). These limitations are overcome by the current work, which also provides a deeper discussion of the results, where the classification of the scenarios and the selection of the parameters are much more elaborate than in~\cite{guijarro2018electronics}.

\section{Model}\label{sec:model}

In this section, we propose a model amenable for the analysis of the service provision by NSTs, within a network slicing framework. The notation used is summarized in Table~\ref{table:notation}.

\begin{table}
\centering
\caption{Summary of notation}
 \renewcommand{\arraystretch}{1.1}
 \begin{tabular}{@{}lp{70mm}@{}}
   \toprule
   Symbol & Meaning \\
   \midrule
   $\mathcal{B}$ 			& Set of cells \\
   $B$ 						& Number of cells \\ 
   $c^{(j)}$				& Capacity of cell~$j$\\  
   $\mathcal{S}$ 			& Set of NSTs \\
   $S$ 						& Number of NSTs \\ 
   $\mathcal{U}$ 			& Set of users \\
   $U$ 						& Number of users \\ 
   $\mathcal{U}^{(j)}$ 		& Set of users in cell $j$\\ 
   $n^{(j)}$				& Number of users in cell $j$\\ 
   $\mathcal{U}_i$ 			& Set of subscribers of NST~$i$\\
   $\mathcal{U}^{(j)}_i$	& Set of subscribers to NST~$i$ in cell~$j$\\
   $n^{(j)}_i$				& Number of subscribers to NST~$i$ in cell~$j$\\ 
   $s_i$					& Share of resources of NST~$i$\\
   $\omega^{(j)}_i$         & Weight assigned by NST~$i$ to the users in $\mathcal{U}^{(j)}_i$\\
   $\vec{w}_i$				& Vector of weights set by NST~$i$\\
   $R^{(j)}_i$ 				& Amount of resources assigned to the users in $\mathcal{U}^{(j)}_i$\\ 
   $v^{(j)}_i$				& Objective part of the utility of a user in  $\mathcal{U}^{(j)}$ subscribing to NST~$i$\\
   $\kappa_{u,i}$			& Unobserved user-specific part of the utility of a user $u$ subscribing to NST~$i$\\
   $r^{(j)}_i$ 				& Amount of resources assigned to each user in $\mathcal{U}^{(j)}_i$\\ 
   $p_i$					& Price paid by the subscribers of NST~$i$\\
   $\alpha$					& User sensitivity parameter\\   
   $\beta$					& User sensitivity parameter, $\beta=\alpha/(\alpha+1)$\\  
   $\suscRatio{j}$			& Subscription ratio in cell $j$\\
   $\ncap{j}$ 				& Normalized capacity of cell $j$\\
   $\rho_i^{(j)}$ 			& Fraction of subscribers in cell $j$ that subscribe to NST~$i$\\
   $\Pi_i$ 					& Revenue of NST~$i$\\
  $\epsilon(\tilde{x})$		& Relative deviation of the approximate value $\tilde{x}$ from the exact value $x$\\
   \bottomrule
 \end{tabular}\label{table:notation}
\end{table}

\subsection{System model}\label{sec:systemmodel}
We focus on mobile service operators and, specifically, on the radio access network.
Then, the network consists of a set of cells $\mathcal{B}=\{1,\ldots,B\}$ and their resources,
where $B$ is the number of cells.
The network resources are managed by an InP and used by a set $\mathcal{S}=\{1,\ldots,S\}$ of NSTs,
where $S$ is the number of NSTs.
The resources of a cell $j \in \mathcal{B}$ are the radio resources of the cell.
The total amount of resources available in cell~$j$ is denoted by~$c^{(j)}$, 
and we refer to it as the \textit{capacity of cell~$j$}.

The resources requested by the NSTs are used to deliver service to a set $\mathcal{U}=\{1,\ldots,U\}$ of users, where $U$ is the number of users.

We define the following subsets of users: 
$\mathcal{U}^{(j)}$, the users in cell $j$;
$\mathcal{U}_i$,  the subscribers of NST~$i$;
and $\mathcal{U}^{(j)}_i=\mathcal{U}^{(j)} \cap \mathcal{U}_i$, the subscribers of NST~$i$ in cell~$j$. 

NST~$i$ is entitled to a share $s_i$ of the total amount of resources available in the network, 
such that $\sum_{i \in \mathcal{S}} s_i= 1$. 
As stated in the introduction, this entitlement may typically result from the operators' decision to pool the networks and to outsource their operation to the InP.

The NST allocates the resource for providing service to its subscribers in the following way. 
NST $i$ distributes its share among its subscribers, assigning a weight $\omega^{(j)}_i$ to $\mathcal{U}^{(j)}_i$, 
such that  $\omega^{(j)}_i > 0$ and $\sum_{j \in \mathcal{B}} \omega^{(j)}_i = s_i \leq 1$.
The weight assignment decision is notified to the InP, who proceeds to perform the actual resource allocation in each individual cell. 
The InP allocates an amount of resources to the set $\mathcal{U}^{(j)}_i$ given by
\begin{equation}\label{eq:rate}
R^{(j)}_i=\frac{\omega^{(j)}_i}{\sum_{t \in \mathcal{S}} \omega^{(j)}_t} c^{(j)}.
\end{equation}
This allocation scheme is known as ``proportional-share algorithm''~\cite{feldman2009, caballero2019}.

Furthermore, in this allocation scheme, NST~$i$ chooses weight $\omega^{(j)}_i$ 
for the set of its subscribers in cell $j$ and the InP performs the actual resource allocation in each individual cell 
according to~\eqref{eq:rate}. 
Proceeding in such indirect way, the capacity constraint in each cell is automatically enforced, i.e., 
$\sum_{i \in \mathcal{S}} R^{(j)}_i = c^{(j)}, \forall j \in \mathcal{B}$.

The rationale of the above allocation rule draws from the Fisher market model. The Fisher market is one of the most fundamental models within mathematical economics~\cite{brainard2005}. The setting is that of a set of buyers ($i=1, \ldots, n$) aiming to purchase quantities ($x_{i,j}$) of multiple goods ($j=1, \ldots, m$), each one under a limited and fixed supply ($\sum_i x_{i,j} \leq c_j$) and given a price system ($p_j$), in a way that maximizes their utility ($u_i(x_{i,1},\ldots, x_{i,m})$) subject to budget constraints ($\sum_j p_j x_{i,j} \leq B_i$). In such a setting, the outcome consisting of a goods allocation and a price system where supply equals demand ($c_j=\sum_i x_{i,j}$) is known as market equilibrium, and has the property that the buyers spend their entire budget.
 A market equilibrium for the Fisher market is guaranteed to exist under mild conditions~\cite{branzei2014}. Note that this market model assumes that prices are taken by the buyers as given, an assumption that is commonly known as \emph{price taking}. It is also known that a market equilibrium is efficient, that is, maximizes the sum of the utilities of the buyers.
When the Fisher market is formulated not in terms of quantities $x_{i,j}$ but in terms of allocated budgets ($b_{i,j}=p_j x_{i,j}$), the supply-equal-to-demand equilibrium condition can be stated as $p_j c_j=\sum b_{i,j}$, which explicitly gives an expression for the prices: $p_j=\sum b_{i,j}/c_j$~\cite{zhang2011}.
The analogy between the Fisher market and our resource allocation model is as follows: the NSTs are the buyers, the resources available at one cell are one good, the NST's share is the buyer's budget, the weight is the part of the budget that each buyer spends on each good, and the utility will be set to the NST's profit. 
Under such an analogy, the allocation rule~\eqref{eq:rate} would be efficient provided that the NSTs were not aware of the whole expression but only of $R^{(j)}_i=\omega^{(j)}_i/p_j$, where the underlay price system would be $p_j=\sum_{t \in \mathcal{S}} \omega^{(j)}_t/c^{(j)}$. This provides a rationale for the allocation rule~\eqref{eq:rate}.
However, in the scenario under study, in which the NSTs are aware of~\eqref{eq:rate}, the setting is no longer price taking but strategic, that is, a game, and the equilibrium allocation will be given not by a market equilibrium, but by a Nash equilibrium, which is typically an inefficient allocation~\cite{branzei2014}, i.e., does not maximize the sum of the utilities of the NSTs. 

\subsection{Economic model}\label{sec:economicmodel}

Each NST provides service to users based on the resource allocation agreed with the InP, according to the description made above. Pricing for the service provision consists of a flat-rate price $p_i$. 
We assume that variable costs incurred by the NSTs are zero, so that only fixed costs are incurred. Furthermore, since the fixed costs are not relevant to the weight decision made by the NSTs, they are not included in the analysis.

In our model, the user choice is the choice of one of the NSTs in $\mathcal{S}$.
We use a discrete-choice model for the modeling of the users' choices, which is frequently used in econometrics~\cite{ben-akiva1985}. Specifically, given a discrete set of options, 
the utility of a user $u \in \mathcal{U}^{(j)}$ making the choice $i$  is assumed to be equal to $v^{(j)}_i + \kappa_{u,i}$: the term $v^{(j)}_i$ encompasses the objective aspects of option $i$ and is the same for all users in $\mathcal{U}^{(j)}_i$, while $\kappa_{u,i}$ is an unobserved user-specific value that is modeled on the global level as a random variable.
From the distribution of these i.i.d. variables, one can compute the probability that a user selects option $i$, and when the user population size is sufficiently large, this is a good approximation of the proportion of users making that choice.

To model the objective part of the users' utility, each subscriber pays the price $p_i$ to NST~$i$, and receives service in cell~$j$ supported by an amount of resources $r^{(j)}_i$. 
Our model focuses primarily on elastic traffic, 
where the users can adapt to the available amount of resources 
and their utility is an increasing and concave function of the amount of resources received.
Such utility function reflects the fact that user satisfaction increases with the amount of resources received,
but exhibits diminishing returns.
Following~\cite{maille2014}, we propose 
\begin{equation}\label{eq:utility}
v^{(j)}_i = \mu \log \left( r^{(j)}_i / p_i\right).
\end{equation}
Firstly,  utility depends logarithmically on the amount of resources, 
as there is increasing evidence that user experience and satisfaction in telecommunication scenarios 
follow logarithmic laws~\cite{reichl2011}. 
Secondly, the dependence on the price is through a negative logarithm, instead; 
or in other words, the ratio $r^{(j)}_i/p_i$ is proposed to be the relevant magnitude for the utility. 
And thirdly, $\mu > 0$ is a sensitivity parameter. 

To model the unobserved user-specific part of the utility, following the literature on discrete-choice models, we assume that each user-specific random variable $\kappa_{u,i}$ follows a Gumbel distribution of mean 0 and parameter $\nu$.
The choice of the Gumbel distribution allows us to obtain a logistic function, as shown below.

With the users' utility modeled as stated above, it can be shown~\cite{train2009} that 
the number of users $n^{(j)}_i$ that subscribe to NST~$i$ over the total number of users in cell~$j$, $n^{(j)}$, is
\begin{equation}\label{eq:users}
 \frac{n^{(j)}_i}{n^{(j)}} = \frac{(r^{(j)}_i/p_i)^\alpha}{\sum_{t \in \mathcal{S}} (r^{(j)}_t/p_t)^\alpha+(r_0^{(j)}/p_0)^\alpha}, 
  \qquad i \in \mathcal{S}, \quad j \in \mathcal{B},
\end{equation}
where $\alpha = \mu/\nu$ is the user sensitivity parameter
(it models the sensitivity to the resources-to-price ratio), 
and $r_0^{(j)}/p_0$ is related to the utility of not subscribing, $v^{(j)}_0$, through~\eqref{eq:utility}:
$r_0^{(j)}/p_0=e^{v^{(j)}_0/\mu}$.
Note that the no-subscription option does not correspond to any network slice in $\mathcal{S}$ and 
therefore no weights are assigned to it. 
The case in which all users in cell~$j$ subscribe 
(i.e., a user in that cell will always be better off by subscribing than by not doing it) 
is captured by letting $v^{(j)}_0 \rightarrow -\infty$ or, equivalently, setting $r_0^{(j)}=0$.
In practice, this case corresponds to a more general situation in which the utility of subscribing to
some of the NSTs clearly outweighs the no-subscription option, that is, 
$r_i^{(j)}/p_i \gg r_0^{(j)}/p_0$ for some $i \in \mathcal{S}$.
Note also that $n^{(j)},\ j \in \mathcal{B}$, are taken as constants, though in a real system they would be random processes due to users mobility.  We assume that, at every cell, entries and exits of users are in balance. Thus, the stationary distribution of the number users in a cell can be modeled as a Poisson random variable (see~\cite{ashtiani2003,zheng2018} for further details).
 The \emph{squared coefficient of variation} of the Poisson distribution is inversely proportional to its mean. Then, if the mean number of users in each cell is moderately high, taking $n^{(j)}$ as a constant equal to the mean number of users in the cell $j$ would be a reasonable simplification.

We assume that the users are price-takers, which is a sensible assumption when there are a sufficiently high number of users in each cell. 

As argued in the next section,  we assume that the service price is the same for every NST: 
\begin{equation}
p_i = p , \quad \forall i \in \mathcal{S}. 
\end{equation}
Besides, without any loss of generality, we set $p_0=1$ to reduce the number of parameters.
The number of subscribers to NST~$i$ in cell~$j$ is then given by
\begin{equation}\label{eq:users1price}
 n^{(j)}_i = n^{(j)} \frac{ (r^{(j)}_i)^\alpha }
 { \sum_{t \in \mathcal{S}}(r^{(j)}_t)^\alpha + (p\,r_0^{(j)})^\alpha }, 
\qquad i \in \mathcal{S}, \quad j \in \mathcal{B}.
\end{equation}

We assume that the resources allocated by the InP to NST~$i$ at a cell are equally shared among the users in 
$\mathcal{U}^{(j)}_i$,
that is, on average, the amount of resources supporting the service to a user is
\begin{equation}\label{eq:rate0price}
r^{(j)}_i=\frac{R^{(j)}_i}{n^{(j)}_i}, 
\qquad i \in \mathcal{S}, \quad j \in \mathcal{B}.
\end{equation}

Taking into account~\eqref{eq:rate}, the amount of resources assigned to a user is
\begin{equation}\label{eq:rate1price}
r^{(j)}_i= \frac{\omega^{(j)}_i}{\sum_{t \in \mathcal{S}} \omega^{(j)}_t} \frac{c^{(j)}}{n^{(j)}_i}, 
\qquad i \in \mathcal{S}, \quad j \in \mathcal{B}
\end{equation}
and substituting~\eqref{eq:rate1price} into~\eqref{eq:users1price}, we come to
\begin{multline}\label{eq:usersFunWeights1}
  \frac{
    \sum_{t \in \mathcal{S}}\left(\frac{\omega^{(j)}_t}{n^{(j)}_t}\right)^{\alpha}
     + 
    \left(\frac{p r_0^{(j)}}{c^{(j)}}  \sum_{t \in \mathcal{S}} \omega^{(j)}_t\right)^{\alpha}
  }
  {n^{(j)}} 
  =
  \frac{(\omega^{(j)}_i)^{\alpha}}{(n^{(j)}_i)^{\alpha+1}},
\\
  i \in \mathcal{S}; \quad j \in \mathcal{B}.
\end{multline}
Let $\penratio{j}$ denote the \textit{subscription ratio} in cell $j$, that is,
the fraction of users in cell $j$ that subscribe to an NST:
\begin{equation}\label{eq:penetration}
\penratio{j} = \frac{1}{n^{(j)}}\sum_{i \in \mathcal{S}} n^{(j)}_i,   \qquad j \in \mathcal{B};
\end{equation}
and let
\begin{equation}\label{eq:normCapacity}
\ncap{j} = \frac{c^{(j)}}{n^{(j)} p\,r_0^{(j)}},   \qquad j \in \mathcal{B}.
\end{equation}
We refer to $\ncap{j}$ as the \textit{normalized capacity} of cell $j$,
and it represents the capacity per monetary unit and per user 
(assuming that the cell capacity is shared equally between all of them)
normalized by the virtual capacity of the no-subscription option, $r_0^{(j)}$.
 
The next proposition
states that the subscription ratio $\penratio{j}$ depends on
the normalized capacity $\ncap{j}$, 
on the user sensitivity $\alpha$ through
\begin{equation}
\beta = \frac{\alpha}{\alpha+1}<1,
\end{equation} 
and on the weights in cell~$j$ of all the NSTs
(i.e., $\omega^{(j)}_i,\ i \in \mathcal{S}$).

\begin{proposition}\label{PROP:PENETRATION}
  \begin{enumerate}
    \item   If $r_0^{(j)}>0$, then the value of $\penratio{j}$ is the unique solution in $(0,1)$ of the equation
      \begin{equation}\label{eq:penetration_NLeq}
      \penratio{j} - \left(\ncap{j}\right)^\beta
                     \frac{\sum_{t \in \mathcal{S}} \left(\omega^{(j)}_t\right)^{\beta}}
                          {\left(\sum_{t \in \mathcal{S}} \omega^{(j)}_t\right)^{\beta}}
                     \left(1-\penratio{j}\right)^{1-\beta} = 0.
      \end{equation}
    \item  If $r_0^{(j)}=0$, then $\penratio{j}=1$.
\end{enumerate}  
\end{proposition}

All the proofs are deferred to the Appendix.

In the general case (i.e., when $r_0^{(j)}>0$),
the previous proposition does not provide a 
closed-form expression for the subscription ratio $\penratio{j}$, 
but a non-linear equation from which it can be obtained numerically.
The following propositions provide some useful insight 
by establishing some properties of $\penratio{j}$ as a function of 
the normalized capacity and the user sensitivity.
Some of these results confirm what intuition suggests.
For example, for given fixed values of the weights,
it seems intuitive that the subscription ratio $\penratio{j}$
is an increasing function of the normalized capacity $\ncap{j}$,
and that a subscription ratio as close as desired to $1$ can be obtained by increasing $\ncap{j}$ sufficiently. 
The impact of the network shares is examined later in Proposition~\ref{PROP:PENETRATION_MINMAX}.

\begin{proposition}\label{PROP:PENETRATION_NORMCAPACITY}
For given fixed values of $\omega^{(j)}_i,\ i \in \mathcal{S}$, the subscription ratio $\sigma^{(j)}$ is an increasing function of $\gamma^{(j)}$ and
\begin{align}
\lim_{\gamma^{(j)}\rightarrow 0} \sigma^{(j)}      &= 0 \\
\lim_{\gamma^{(j)}\rightarrow \infty} \sigma^{(j)} &= 1.	
\end{align}
\end{proposition}

\begin{proposition}\label{PROP:PENETRATION_SENSITIVITY}
\begin{align}
  \lim_{\beta\rightarrow 0} \sigma^{(j)} &= \frac{\cardS}{1+\cardS}\\
  \lim_{\beta\rightarrow 1} \sigma^{(j)} &= \min\big(1, \gamma^{(j)}\big).
\end{align}
\end{proposition}

Let $\rho_i^{(j)}$ denote the \emph{fraction of subscribers} in cell $j$ that subscribe to NST~$i$:
\begin{equation}\label{eq:rho_ij}
	\rho_i^{(j)} = \frac{n^{(j)}_i}{\sum_{t \in \mathcal{S}} n^{(j)}_t}
	             = \frac{n^{(j)}_i}{\penratio{j} n^{(j)}}.
\end{equation}
This fraction can be expressed as a function of the weights at that cell
as given in the following proposition.  

\begin{proposition}\label{PROP:N_RS}
  For each cell $j \in \mathcal{B}$ and  each NST $i \in \mathcal{S}$
  \begin{equation}\label{eq:n_rs}
	 \rho_i^{(j)} =  
	         \frac{(\omega^{(j)}_i)^\beta}{\sum_{t \in \mathcal{S}}(\omega^{(j)}_t)^\beta}.
  \end{equation}
\end{proposition}


\section{Game model and analysis}\label{sec:analysis}

The revenue of NST~$i$ is equal to the total amount charged to its subscribers, that is,
\begin{equation}\label{eq:profits}
\Pi_i = 
   p \sum_{j \in \mathcal{B}} n^{(j)}_i, 
  \qquad i \in \mathcal{S}.
\end{equation} 
Using~\eqref{eq:rho_ij} we can express the revenue as follows:
\begin{equation}\label{eq:profitsweightsgeneral}
\Pi_i = p\sum_{j \in \mathcal{B}} 
                  n^{(j)}\penratio{j} 
				  \rho_i^{(j)},
    \qquad i \in \mathcal{S},
\end{equation}
which depends not only on the weights $\omega^{(j)}_i$ set by NST~$i$, but also on the weights set by the other NSTs.
Each NST is assumed to operate in order to maximize its revenues. 

We assume that the competition is not in terms of prices. 
Instead, we analyze the competition between the NSTs in terms of quality of service.
A scenario where the user subscription is driven by the currently received quality of service and not by the price is not uncommon (see, e.g., \cite{allon2007}). We provide three alternative rationales for this situation:
\begin{enumerate}

\item That a regulatory authority is enforcing price control over the provision of the service, i.e., the service price is fixed by the authority and therefore not under the control of the operators/tenants.

\item That the service price is agreed on a long-term contract, e.g., as part of a bundled offer.

\item That the operators/tenants are wholesale roaming service providers, i.e., that they provide the service to roaming users. We provide three examples of this scenario:

\begin{itemize}

\item An example currently in exploitation is a mobile communication service. The users pay their home operator according to their retail contract; the home operator signs roaming contracts with several local operators; and the local operators request the resources from the InP-operated pool and compete to provide service to the roaming users. The roaming contracts will fix the revenue per user served by a local operator, so that the local operators can only compete in QoS.

\item A near-future scenario would be a roadside assistance service. The drivers would pay their car manufacturer for the service; the car manufacturer would sign contracts with several local assistance service providers (each one having a partnership agreement with a local operator). And the local service provider/operators would request the resources from the InP-operated pool and compete to provide service to the drivers. 

\item An advertiser-funded streaming service. Visitors to shopping malls would be willing to stream advertiser-funded content provided they are not consuming their data cap. The content provider would sign contracts with local micro-operators having deployed WiFi infrastructure in the premises, each provider committing to pay a tariff per connected shopper. Each micro-operator would request the resources from the InP-operated pool and compete to provide service to the shoppers. 

\end{itemize}

\end{enumerate}

The analysis is focused, therefore, on how each NST sets the weight $\omega^{(j)}_i$ in cell $j$ in order to attract as many users as possible.~
The vector of weights set by NST~$i$, with one component $\omega^{(j)}_i$ for each $j \in \mathcal{B}$, 
is denoted by  $\vec{w}_i \in (0,1)^{\cardB}$.

To decide on its strategy NST~$i$ has to solve the following revenue maximization problem:
\begin{equation} \label{eq:bestRespDef}
\begin{aligned} 
	& \underset{\vec{w}_i}{\text{max }} & & \Pi_i (\vec{w}_i, \vec{w}_{-i}) \\
	& \text{subject to} & & \sum_{j \in \mathcal{B}} \omega^{(j)}_i \leq s_i\\ 
	& & & \vec{w}_i \in (0,1)^{\cardB}, %
\end{aligned} 
\end{equation}
where $-i$ refers to all NSTs other than NST~$i$ and, hence, $\vec{w}_{-i} \in (0,1)^{(\cardS-1)\cardB}$.

As shown above, there is a strategic dependence of the revenue of NST~$i$ on the weights set by the competing NSTs. 
This fact allows us to model the combined revenue maximization problems as a strategic game.
We will use a simultaneous one-shot game model for the analysis. 
The solution for the game, i.e., a strategy for each NST facing the interrelated maximization problems stated in~\eqref{eq:bestRespDef}, is the Nash equilibrium.
In the Nash equilibrium, the weights that each NST chooses in each cell are such that it gets no revenue improvement 
from changing the weights assuming that the competitor NSTs do not deviate from the equilibrium weights. 
Note that the Nash equilibrium does not provide by itself any prescription for the competing NSTs about how to reach the equilibrium.
Note that the Nash equilibrium does not provide by itself any prescription for the competing NSTs about how to reach the equilibrium, and that this it is not necessarily optimal. We have postponed for future work the quantification of its optimality in terms of the price of anarchy.

Let $\vec{B}_i(\vec{w}_{-i})$ be the best response function for NST~$i$,
which assigns the solution of \eqref{eq:bestRespDef} to each $\vec{w}_{-i}$.
If ${\vec{w}^*_i}_{i \in \mathcal{S}}$ is a Nash equilibrium,
then $\vec{w}^*_i \in \vec{B}_i(\vec{w}^*_{-i})$ for all $i \in \mathcal{S}$.
Now, to find the Nash equilibrium to our problem
we try to solve the equation  
$\vec{w}_i  = \vec{B}_i(\vec{w}_{-i})$ for a generic $i \in \mathcal{S}$,
so that it holds for all NSTs simultaneously.


We propose the following form for the Nash equilibrium,
which we refer to as the \textit{proposed solution}:
\begin{equation}\label{eq:genSolWeights}
\omega^{(j)}_i =\frac{\penratio{j} n^{(j)}}{\sum_{k \in \mathcal{B}} \penratio{k} n^{(k)}} s_i,
\qquad i \in \mathcal{S},  \quad j \in \mathcal{B}.
\end{equation}
In this solution, each NST chooses, for each cell, a weight equal to 
the product of its share by 
the ratio between the total number of subscribers in the cell and the total number of subscribers in the network. 
It should be noted that: 
(i) Eq.~\eqref{eq:genSolWeights} does not provide by itself a solution for the weights in the equilibrium,
since in this expression the weights depend on the subscription ratios, which in turn depend on the weights 
(see Proposition~\ref{PROP:PENETRATION});
(ii) the solution that can be derived from~\eqref{eq:genSolWeights} is exact 
(i.e., it is exactly equal to the Nash equilibrium) 
when the normalized capacities of the cells satisfy certain requirements (to be specified below),
and it is a very accurate approximation otherwise.

Next, we address these two issues. 
First we show how the form of the solution given in~\eqref{eq:genSolWeights} 
can be used to obtain the actual solution and study the properties of the solution.
Then, we present the condition under which the proposed solution is exact, 
and provide the mathematical arguments that lead to this result.
Finally, in Section~\ref{sec:results} 
an extensive set of numerical experiments is used to show that, 
when the proposed solution is not exact, it provides a highly accurate approximation.

Substituting~\eqref{eq:genSolWeights} into~\eqref{eq:penetration_NLeq} 
we obtain
\begin{equation}\label{eq:penetration_NLeq_ref}
\penratio{j} - \left(\ncap{j}\right)^\beta
               \sum_{t \in \mathcal{S}} s_t^{\beta}
               \left(1-\penratio{j}\right)^{1-\beta} = 0,
\end{equation}
from where the value of $\penratio{j}$ can be obtained numerically, 
with low complexity and in a scalable way.  Also,  $\penratio{j}$ solely depends on the normalized capacity of the cell and on the share of all NSTs, so that each tenant could calculate its weights for the proposed solution without knowing the weights chosen by the others, which preserves privacy of their business strategies.
We observe that $\penratio{j}$ solely depends on the normalized capacity of the cell 
and on the share of all NSTs.
Similarly, substituting~\eqref{eq:genSolWeights} into~\eqref{eq:n_rs} yields
   \begin{equation}\label{eq:n_rs_ref}
	   \rho_i^{(j)} =  \rho_i,
   \end{equation}
where	  
   \begin{equation}\label{eq:rho_i}
	   \rho_i =
	    \frac{s_i^\beta}{\sum_{t \in \mathcal{S}}s_t^\beta},
	       \qquad i \in \mathcal{S}.
   \end{equation}
This tells us that, in each cell, 
the subscribers to the service are split proportionally between the NSTs 
with the coefficients $s_i^\beta$.
We observe that $\rho_i^{(j)}=\rho_i$ solely depends on the shares of the NSTs and on $\beta$, and, consequently,
it is the same across all cells.
From these results, each NST can calculate its weights corresponding to the proposed solution by solving~\eqref{eq:penetration_NLeq_ref} and then applying~\eqref{eq:genSolWeights}. For this, the only information required is the normalized capacity in all cells, the number of users in all cells, and the share of all tenants. 

\begin{figure*}[!t]
  \normalsize
  \setcounter{MYtempeqncnt}{\value{equation}}
  \setcounter{equation}{39}
   \begin{multline}\label{eq:Djj}
     \frac{\partial^2 \Pi_i}{\partial \big(\omega^{(j)}_i\big)^2 } =
	    \frac{p\beta n^{(j)}\penratio{j}\rho_i^{(j)}}
		     {\big(\omega^{(j)}_i\big)^2 }	
		\Bigg(
		  \frac{1-\penratio{j}}{1-\beta\penratio{j}} 
		  \bigg(	
		     \Big(\rho_i^{(j)} - x^{(j)}_i\Big)^2  
			 \frac{\beta}{1-\beta\penratio{j}} 
			 \left(\frac{1-\penratio{j}}{1-\beta\penratio{j}} - \penratio{j}\right)
			 + \Big(x^{(j)}_i \Big)^2
			 \\
			 + \beta\Big(1-\rho_i^{(j)}\Big)\Big(3\rho_i^{(j)}-2 x^{(j)}_i\Big)
			 - \rho_i^{(j)} 
		  \bigg)
		  -\Big(1-\rho_i^{(j)}\Big)\Big(1-\beta+2\beta\rho_i^{(j)}\Big)	
		\Bigg)	
   \end{multline}
  \setcounter{equation}{\value{MYtempeqncnt}}
  \hrulefill
\end{figure*}

As a complementary result to that of Proposition~\ref{PROP:PENETRATION_NORMCAPACITY},
the following proposition establishes that in the equilibrium the subscription ratio 
in all cells is maximum if all NSTs have the same share.

\begin{proposition}\label{PROP:PENETRATION_MINMAX}
For given fixed values of the normalized capacities $\ncap{j},\ j \in \mathcal{B}$,
the subscription ratio in each cell 
is maximum when all NSTs are allocated the same share
   \begin{equation}\label{eq:equalShares}
     s_i=\frac{1}{\cardS},  \qquad i \in \mathcal{S}
   \end{equation}
and is minimum when a single NSTs is allocated all resources
   \begin{equation}\label{eq:unEqualShares}
     s_i=1,  \quad i \in \mathcal{S} 
	 \qquad \text{and} \qquad
	 s_t=0,  \quad t \in \mathcal{S}\setminus\{i\}.
   \end{equation}
In other words, if
\begin{align}\label{eq:penetration_Max}
	\penratio{j}_{\max} &= \max\left\{\penratio{j} : s_i \in [0,1], \ i \in \mathcal{S};
	                                           \quad\sum_{i \in \mathcal{S}} s_t=1 \right\} \\
	\penratio{j}_{\min} &= \min\left\{\penratio{j} : s_i \in [0,1], \ i \in \mathcal{S};
	                                           \quad\sum_{i \in \mathcal{S}} s_t=1 \right\}				
\end{align}
$\penratio{j}_{\max}$ and $\penratio{j}_{\min}$ can be obtained as the solutions to
\begin{align}
&\penratio{j}_{\max} - \cardS^{1-\beta}\left(\ncap{j}\right)^\beta
               \left(1-\penratio{j}_{\max}\right)^{1-\beta} = 0
\\			
&\penratio{j}_{\min} - \left(\ncap{j}\right)^\beta
               \left(1-\penratio{j}_{\min}\right)^{1-\beta} = 0.			
\end{align}

\end{proposition}


When all cells have the same normalized capacity, 
from~\eqref{eq:penetration_NLeq_ref} we see that the subscription ratio is the same for all cells.
Furthermore, when $r^{(j)}_0 = 0$ we have $\penratio{j}=1$ (Proposition~\ref{PROP:PENETRATION}).
Thus, if $r^{(j)}_0 = 0, \ j \in \mathcal{B}$, the subscription ratio is also the same for all cells.
In practice, the latter case captures those scenarios in which the normalized capacity of all cells
is sufficiently high, so that all (or nearly all) users subscribe to an NST.

These observations are summarized in the next proposition, 
which also draws additional conclusions.

\begin{proposition}\label{PROP:EQUALPENETRATIONS}
If one of the two following conditions is satisfied:
\begin{itemize}
	\item $r^{(j)}_0 = 0, \quad j \in \mathcal{B}$
	\item $\ncap{j} = \gamma, \quad j \in \mathcal{B}$
\end{itemize}
then:
\begin{enumerate}
	\item The  subscription ratio is the same in all cells:
           \begin{equation}\label{eq:penetration_ref}
               \penratio{j} = \sigma, \quad j \in \mathcal{B}.
           \end{equation}	
	\item The weights of the proposed solution become
           \begin{equation}\label{eq:solWeights_ref}
           \omega^{(j)}_i =\frac{n^{(j)}}{n}s_i,
                 \qquad i \in \mathcal{S},  \quad j \in \mathcal{B}.
           \end{equation}
\end{enumerate}
\end{proposition}

We would like to note that~\eqref{eq:solWeights_ref}, which shows that the equilibrium strategy for an NST is to distribute its share proportionally to the number of users in each cell, is not the result of a centralized decision made by the InP, but the result of a game where each NST acts selfishly.

Under the condition of Proposition~\ref{PROP:EQUALPENETRATIONS} the solution to the Nash equilibrium problem
given by~\eqref{eq:solWeights_ref} is exact and all results 
derived from the proposed solution
hold exactly.
If none of the conditions of Proposition~\ref{PROP:EQUALPENETRATIONS} is met, 
then the proposed solution and the rest of the results are approximate only, although
with high accuracy as will be shown in Section~\ref{sec:results}.

In the remainder of this section we discuss the arguments that support the exactness,
and also the uniqueness, of the proposed solution when the required conditions are met.


We derive the solution to the revenue maximization problem of NST~$i$, 
which will yield the best response function $\vec{B}_i(\vec{w}_{-i})$. 
The Karush-Kuhn-Tucker conditions (KKT) for NST~$i$ are as follows:
\begin{align} 
    \label{eq:stationary}
    &\nabla\Pi_i(\vec{w}_i) = 
              \mu_i \nabla \left(\sum_{j \in \mathcal{B}} \omega^{(j)}_i - s_i\right)     
    \\
	\label{eq:posMultiplier}
	&\mu_i \geq 0, 
	\\
	\label{eq:feasibility}
	&\sum_{j \in \mathcal{B}} \omega^{(j)}_i - s_i \leq 0, 
	\\
	\label{eq:complementarySlackness}
	&\mu_i\left(\sum_{j \in \mathcal{B}} \omega^{(j)}_i- s_i\right)=0.
\end{align}

The next proposition gives expressions for the first and second derivatives of 
$\Pi_i$ with respect to the weights of NST~$i$
\begin{proposition}\label{PROP:DERIVATIVES}
  For $j,k \in \mathcal{B}$, $j\neq k$, we have
  \begin{multline} \label{eq:Dj}
	 \frac{\partial   \Pi_i}{\partial \omega^{(j)}_i}	= 
	    \frac{p\beta n^{(j)}\penratio{j}\rho_i^{(j)}}
		     {\omega^{(j)}_i (1-\beta\penratio{j}) }
		\\ 
		\times	
		\left(
		  (1-\beta)(1-\rho_i^{(j)})\penratio{j} 
		  + 
		  (1-x^{(j)}_i)(1-\penratio{j}) 
		\right),
  \end{multline}		
   \begin{equation}\label{eq:Djk}
     \frac{\partial^2 \Pi_i}{\partial \omega^{(j)}_i \partial \omega^{(k)}_i} =0
   \end{equation}
   %
   \addtocounter{equation}{1}
   and~\eqref{eq:Djj} at the top of the page,
   where 
   \begin{equation} \label{eq:xij}
   	 x^{(j)}_i = \frac{\omega^{(j)}_i }{\sum_{t\in \mathcal{S}}\omega^{(j)}_t}.
   \end{equation}
\end{proposition}

The condition~\eqref{eq:stationary} can be simplified as 
\begin{equation} \label{eq:br2}
	\Dparcial{\Pi_i}{\omega^{(j)}_i}	= \mu_i, \qquad j \in \mathcal{B}.
\end{equation}
Now, from~\eqref{eq:Dj} it is immediate that 
$ \mu_i=\partial\Pi_i/\partial\omega^{(j)}_i > 0$,
which combined with~\eqref{eq:complementarySlackness} shows that
the constraint of~\eqref{eq:feasibility} must be active for all optimal points:
\begin{equation}\label{eq:equality}
	\sum_{j \in \mathcal{B}} \omega^{(j)}_i = s_i. 
\end{equation}

The next proposition states that when the subscription ratio is the same across all cells,
the proposed solution, which in that case takes the form given in~\eqref{eq:solWeights_ref}, 
satisfies the necessary KKT conditions.

\begin{proposition}\label{PROP:PSKKTNECCOND}
Under the condition of Proposition~\ref{PROP:EQUALPENETRATIONS},
the weights of~\eqref{eq:solWeights_ref} and the multiplier 
\begin{equation}\label{eq:multiplier}
	\mu_i = \frac{p\beta n\sigma\rho_i}{s_i (1-\beta\sigma) }	
		    \left( (1-\beta)(1-\rho_i)\sigma + (1-s_i)(1-\sigma) \right),
\end{equation}
satisfy the conditions~\eqref{eq:stationary}--\eqref{eq:complementarySlackness}.
\end{proposition}

From the above result, we can only conclude that the proposed solution meets the necessary conditions
to be a maximum point.
Nevertheless,
from~\eqref{eq:Djj} it can be seen that
\begin{equation}\label{eq:lim_Djj}
 \lim_{\penratio{j}\rightarrow 1} \frac{\partial^2 \Pi_i}{\partial \big(\omega^{(j)}_i\big)^2 } =
	    -\frac{p\beta n^{(j)}\rho_i^{(j)}}
		     {\big(\omega^{(j)}_i\big)^2 }	
		  \Big(1-\rho_i^{(j)}\Big)\Big(1-\beta+2\beta\rho_i^{(j)}\Big)	
		< 0.
\end{equation}
Thus, if the subscription ratios are sufficiently close to 1, we
have that $\partial^2 \Pi_i/\partial \big(\omega^{(j)}_i\big)^2 <0$,
which along with~\eqref{eq:Djk} implies the strict convexity of the maximization problem of~\eqref{eq:bestRespDef}.
In this case, the KKT conditions are also sufficient,
and if there is a maximum point, it is unique.

We recall that the subscription ratio in a cell is high (i.e., close to $1$) in the following cases:
\begin{itemize}
	\item The normalized capacity of the cell is sufficiently high, 
	      so that nearly all users subscribe to an NST 
		  (Propositions~\ref{PROP:PENETRATION} and~\ref{PROP:PENETRATION_NORMCAPACITY}). 
    \item The user sensitivity is low and the number of NSTs is high 
		  (Proposition~\ref{PROP:PENETRATION_SENSITIVITY}).
	\item The user sensitivity is high and the normalized capacity is close to $1$ or higher 
		  (Proposition~\ref{PROP:PENETRATION_SENSITIVITY}).
\end{itemize}	

When none of the conditions above is satisfied we are not able to provide  mathematical guarantees on the solution.
To validate the properties of the proposed solution under more general conditions, 
we have conducted an extensive set of numerical experiments under a wide range of conditions, 
as reported in the next section.
In all our experiments the numerical solution was equal to the proposed solution,
when the condition of Proposition~\ref{PROP:EQUALPENETRATIONS} was satisfied,
and very close to the proposed solution otherwise.


\section{Setup of the numerical experiments}\label{sec:setup}

In our numerical experiments  we distinguish two cases, which we refer to as 
\emph{homogeneous cells} and \emph{heterogeneous cells}. 
In the homogeneous cells case the condition of Proposition~\ref{PROP:EQUALPENETRATIONS} is satisfied
 (and hence the subscription ratio is the same across all cells),
while the heterogeneous cells case is the general one.
As stated in Section~\ref{sec:analysis}, 
if cells are homogeneous the proposed solution is an exact solution,
while otherwise it is an approximation. 
In the next section we discuss the properties of the proposed solution when cells are homogeneous, 
and establish the accuracy of the approximation by quantifying its deviation from the equilibrium 
when cells are heterogeneous.

For the proposed solution,  the subscription ratios $\sigma^{(j)}$ at the equilibrium are obtained 
by solving numerically~\eqref{eq:penetration_NLeq_ref}, 
and the fractions of subscribing users $\rho_i$ at the equilibrium have been obtained 
by applying~\eqref{eq:rho_i}. This has been done in both cases: homogeneous cells and heterogeneous cells.

Additionally, in the two cases mentioned, the game has also been solved numerically by means of 
asynchronous best-response  dynamics (ABRD). 
Note that, in this work, the ABRD is borrowed as an algorithm for the numerical computation of the Nash equilibrium, 
not as a model for the dynamic behavior of the NSTs.
In our ABRD implementation, starting from  $w_i^{(r)}=s_i/\cardB$,  the weights are recalculated in repeated steps until they converge to almost fixed values. 
In each step, NSTs are ordered randomly and each of them calculates sequentially its best response~\eqref{eq:bestRespDef}.
Best responses are calculated numerically by means of a heuristic optimization algorithm based on the cuckoo search algorithm~\cite{walton2011}.
This is a gradient-free algorithm with a complexity that grows exponentially with $\cardB$ and $\cardS$, but in these experiments it has turned out that ABRD does not require a high accuracy in the best response. Besides, ABRD convergence is fast: from an initial state, only about 10 iterations are required.

The pseudocode of ABRD procedure is:
\begin{algorithmic}
\Procedure{ABRD}{$\mathcal{S}$}
\For{$i\in\mathcal{S}$}
\For{$r\in\mathcal{B}$}
$w_i^{(r)}=s_i/\cardB$
\EndFor
\EndFor
\Repeat
\State randomize order of elements in $\mathcal{S}$
\For{$i\in\mathcal{S}$}
\State $\vec{w}_i=\vec{B}_i(\vec{w}_{-i})$   
\EndFor
\Until{variations of $\vec{w}_i, \forall i\in\mathcal{S}$ are under a threshold}
\EndProcedure
\end{algorithmic}

We have verified that ABRD converged in all cases, and that, in the homogeneous cells case, 
the results obtained matched exactly with those given by the proposed solution.
On the other hand, the results obtained through ABRD in the heterogeneous cells case 
have been compared with those of the proposed solution 
to evaluate the accuracy of the approximation given by the latter.

When cells are heterogeneous the proposed solution does not provide the exact equilibrium.
In the numerical experiments for the heterogeneous cells case, 
to distinguish the proposed solution from the exact one,
we denote by $\tilde{w}_i^{(r)}$,  $\tilde{\rho}_i$ and  $\tilde{\sigma}^{(r)}$, 
respectively, 
the weights, fractions of subscribers and subscription ratios at the equilibrium, computed from the proposed solution; 
and keep the original notation ($w_i^{(r)}$,  $\rho_i^{(r)}$ and  $\sigma^{(r)}$) for the results obtained with ABRD. 

To verify the accuracy of the proposed solution, in the next Section,
its deviation from the exact solution is assessed over a wide range of the system parameters.
For doing so, a  diversity of scenario types is considered,
each type with a different number of NSTs, number of cells, value of $\alpha$ and range of $\gamma^{(r)}$.
For each type of scenario,  $1000$ scenarios are solved with both the proposed solution and  ABRD. 
In each scenario, each $\gamma^{(r)},\; r\in\mathcal{B}$,
has been generated randomly with uniform distribution in the range $[\gamma_{\min},\gamma_{\max}]$, 
and the shares have been generated as an equiprobable random vector in the 
$(\cardS-1)$-simplex 
$\{(s_1\dots s_{\cardS})\in\mathbb{R}^{\cardS}\mid \sum_{i\in\mathcal{S}}s_i=1,\; s_i\geq 0.1 \}$.
The relative deviations of the subscription ratio, $\epsilon(\tilde{\sigma}^{(r)})$, 
and of the fraction of subscribers, $\epsilon(\tilde{\rho}_i)$, 
are registered for all scenarios.
They are calculated as follows:
$\epsilon(\tilde{\sigma}^{(r)})=(\tilde{\sigma}^{(r)}-\sigma^{(r)})/\sigma^{(r)}$
and $\epsilon(\tilde{\rho}_i)=(\tilde{\rho}_i-\rho_i^{(r)})/\rho_i^{(r)}$.

\section{Results and discussion}\label{sec:results}
 
In this section we discuss the properties of the proposed solution when cells are homogeneous, and establish the accuracy of the approximation by quantifying its deviation from the equilibrium when cells are heterogeneous.

\subsection{Homogeneous cells}\label{sec:homogeneous}
When conditions of Proposition~\ref{PROP:EQUALPENETRATIONS} are satisfied, there is a unique value of $\gamma$, and
\begin{itemize}
\item the subscription ratio is the same in all cells and can be obtained from~\eqref{eq:penetration_NLeq_ref},
making  $\sigma^{(r)}=\sigma$, and depends on $\gamma$, on the shares and on $\alpha$,
\item and the exact fractions of subscribers are given by~\eqref{eq:rho_i}, and depend on the shares and on $\alpha$, but not on $\gamma$.
\end{itemize} 

We first discuss the results for a scenario where all NSTs have the same share ($s_i=1/\cardS,\; i\in \mathcal{S}$). 
In this case, the fraction of subscribers of each NST is $1/\cardS$, and the subscription ratio is $\sigma=\sigma^{(j)}_{\max}$ given by~\eqref{eq:penetration_Max}.

\begin{figure}[t]
\begin{center}
\includegraphics[width=\columnwidth]{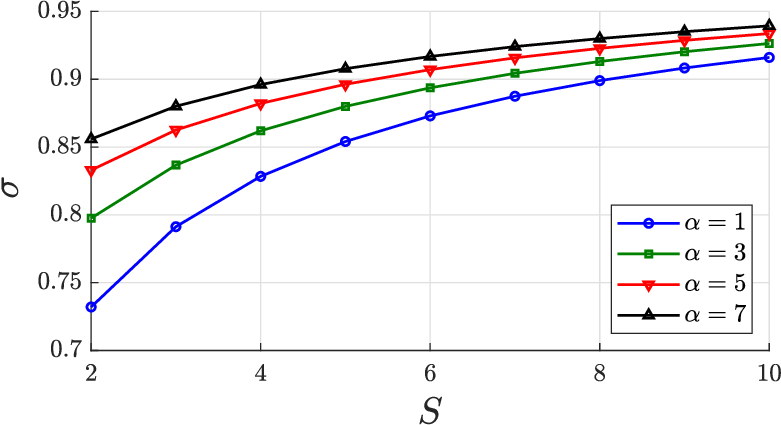}
\caption{Subscription ratio at the equilibrium as a function of the number of NSTs for different values of $\alpha$.}
\label{fig:symSalfa}
\end{center}
\end{figure}

\begin{figure}
\begin{center}
\includegraphics[width=\columnwidth]{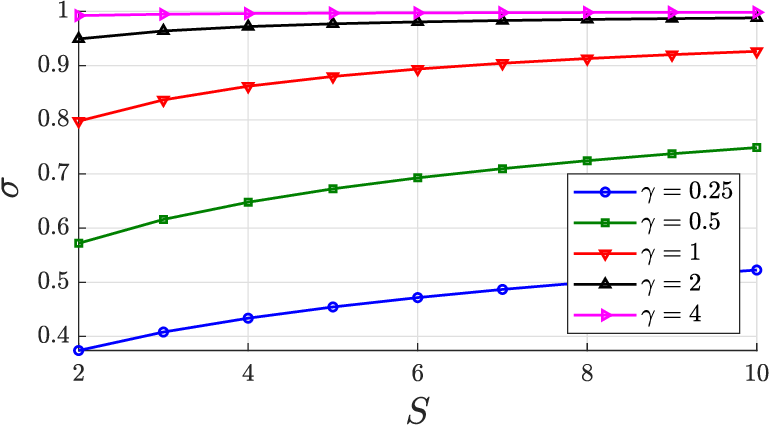}
\caption{Subscription ratio at the equilibrium as a function of the number of NSTs for different values of $\gamma$.}
\label{fig:symSgamma}
\end{center}
\end{figure}

In Figs.~\ref{fig:symSalfa} and~\ref{fig:symSgamma}, the subscription ratio is represented as a function of the number of NSTs, 
for $\gamma=1$ and different values of $\alpha$ (Fig.~\ref{fig:symSalfa}) and for $\alpha=3$ and different values of $\gamma$ (Fig.~\ref{fig:symSgamma}).
In all cases, as the number of NSTs increases, the subscription ratio increases, as can be derived from~\eqref{eq:penetration_Max} (note that $1-\beta>0$).
This can be interpreted as that an increase in the diversity of the service offering increases the subscription ratio. 
Fig.~\ref{fig:symSalfa} also shows that the subscription ratio increases with the user sensitivity. 
Fig.~\ref{fig:symSgamma} also shows, in accordance with Proposition~\ref{PROP:PENETRATION_NORMCAPACITY}, that the greater the normalized capacity, the greater the subscription ratio. 
For $\gamma=4$, the subscription ratio is close to 1, 
which means that if the normalized capacity is high enough,
the result is practically the same as for $r_0^{(r)}=0$ in all cells.

We now investigate how the asymmetry between the NSTs in terms of share affects the results. 
For this, we illustrate a scenario with 4 NSTs and $\alpha=3$, and show the results obtained as a function of the share of one of them.

\begin{figure}
\begin{center}
\includegraphics[width=\columnwidth]{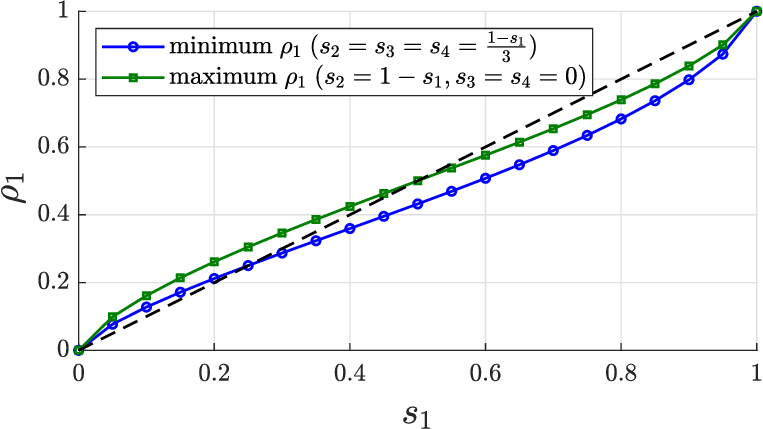}
\caption{Maximum and minimum NST~1's fraction of subscribers at the equilibrium as a function of its share.}\label{fig:asym1n}
\end{center}
\end{figure}

Fig.~\ref{fig:asym1n} shows NST~1's fraction of subscribers at the equilibrium as a function of its share in two different situations: 
one in which the remaining share is shared equally between the rest of NSTs ($s_2=s_3=s_4=(1-s_1)/3$) and another one in which one NST keeps all the remaining share ($s_2=1-s_1,\; s_3=s_4=0$). 
From~\eqref{eq:rho_i}, it is easy to show that the second situation (the one in which the remaining share is distributed as unequally as possible) is the most favorable for NST~1; thus, the corresponding curve represents the maximum fraction of subscribers that NST 1 can obtain. 
It can also be checked that the first situation, in which the remaining share is distributed equally, is the most unfavorable for NST~1, and now the curve represents the minimum fraction of subscribers. 
For any other distribution of the remaining share, the curve would run between these two bounds.

\begin{figure}
\begin{center}
\includegraphics[width=\columnwidth]{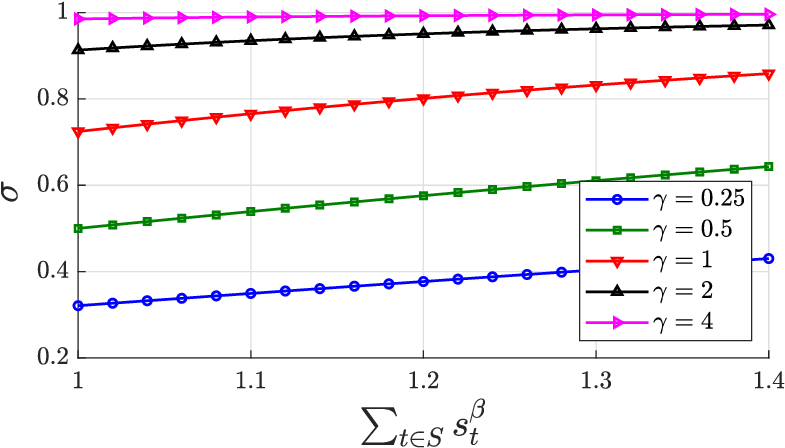}
\caption{Subscription ratio at the equilibrium as a function of the term $\sum_{t\in S}s_t^\beta$ (share equality) for different values of $\gamma$.}
\label{fig:asym1pSumS}
\end{center}
\end{figure}

Fig.~\ref{fig:asym1pSumS} shows the subscription ratio as a function of the factor $\sum_{t\in \mathcal{S}} s_t^\beta$ for several values of $\gamma$. 
This factor depends on the degree of share equality and, together with $\gamma^\beta$, determines the result of~\eqref{eq:penetration_NLeq_ref}. 
The figure shows that the higher the normalized capacity and the share equality, 
the higher the subscription ratio (Proposition~\ref{PROP:PENETRATION_NORMCAPACITY}).
Also, according to Proposition~\ref{PROP:PENETRATION_MINMAX}, the maximum values of $\sigma$ correspond to $s_i=1/4$, when all shares are equal, and the minimum values are  
when a single NST takes all the share ($s_1=1,\;s_2=s_3=s_4=0$).


\subsection{Heterogeneous cells}
 
When cells are heterogeneous the proposed solution does not provide the exact equilibrium, but here we show that it provides an accurate approximation.
Note that, although  NST~$i$'s fraction of subscribers obtained with the proposed solution ($\tilde{\rho}_i$) is the same for all cells, 
the one obtained with ABRD ($\rho_i^{(r)}$) is not.
Besides, 
neither of the two subscription ratios ($\sigma^{(r)}$ or $\tilde{\sigma}^{(r)}$) is the same for all cells.

\begin{figure}
\begin{center}
\includegraphics[width=\columnwidth]{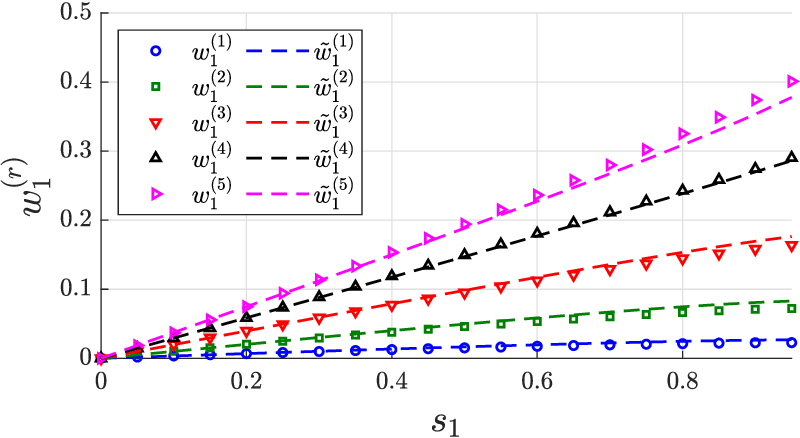}
\caption{NST 1's weights at the equilibrium as a function of NST~1's share, compared with the proposed solution.}
\label{fig:hetw1a}
\end{center}
\end{figure}

\begin{figure}
\begin{center}
\includegraphics[width=\columnwidth]{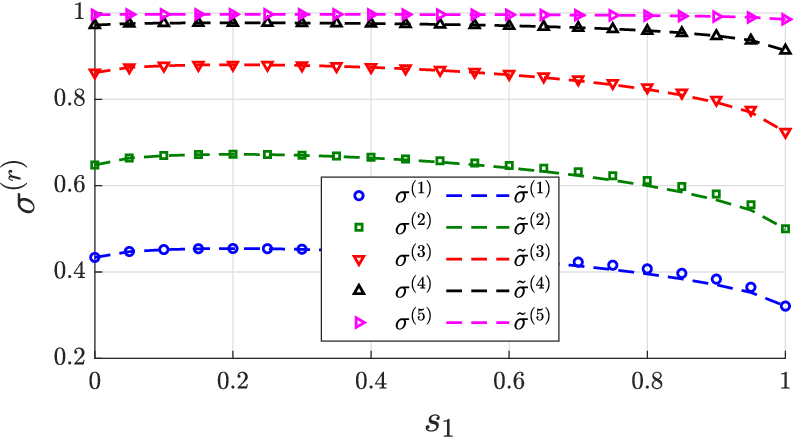}
\caption{Subscription ratios in each cell at the equilibrium as a function of NST~1's share, compared with the proposed solution.}\label{fig:hetp}
\end{center}
\end{figure}

\begin{figure}
\begin{center}
\includegraphics[width=\columnwidth]{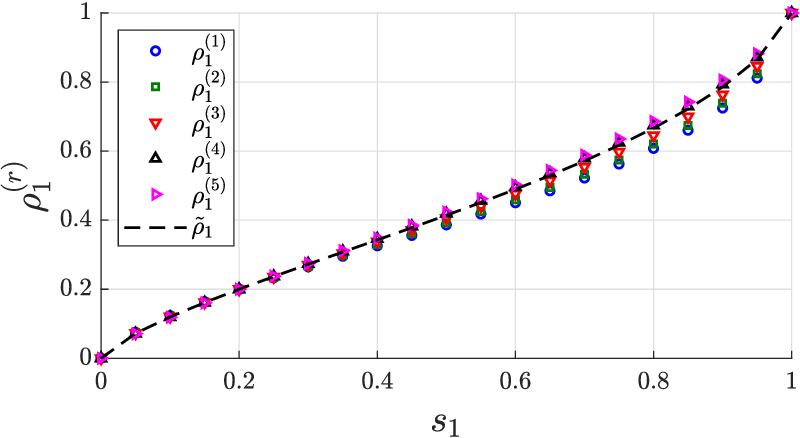}
\caption{NST~1's fraction of subscribers in each cell at the equilibrium as a function of NST~1's share, compared with the proposed solution.}\label{fig:het1n1}
\end{center}
\end{figure}

The plots in Figs.~\ref{fig:hetw1a}--\ref{fig:het1n1} correspond to a scenario with 4~NSTs, 5 cells and $\alpha=3$. 
NST~1's share ranges from $0$ to $1$ and the remaining share is equally distributed between the rest of NSTs.
The numbers of users are $(n^{(1)},\dots, n^{(5)})=(100,200,300,400,500)$, 
and the values of $c^{(r)}$ and $r_0^{(r)}$ have been chosen so that $(\gamma^{(1)},\dots,\gamma^{(5)})=(0.25,0.5,1,2,4)$. 
The marks represent the values obtained with ABRD, 
while the dashed lines represent the results of the proposed solution.

As seen in Figs.~\ref{fig:hetw1a}, NST~1's weights obtained with ABRD are very close to the proposed solution. 
In the cells with lower $\gamma^{(r)}$ (cells 1, 2 and 3), they are slightly below the proposed solution, 
while in the cells with higher $\gamma^{(r)}$ (cells 4 and 5) they are slightly above.
The subscription ratios in each cell are represented in Fig.~\ref{fig:hetp},
where the values obtained with both methods can hardly be distinguished from each other. 
Finally, in Fig.~\ref{fig:het1n1} it can be seen that NST~1's fractions of subscribers are almost the same at all the cells,
and that they are very close to the result of the proposed solution, 
which in fact is the same for all cells.
Just like with the weights, the fractions of subscribers obtained with ABRD are slightly lower than those of the proposed solution in the cells with lower $\gamma^{(r)}$, 
and slightly higher in the cells with higher $\gamma^{(r)}$.

\begin{figure}
\begin{center}
\includegraphics[width=\columnwidth]{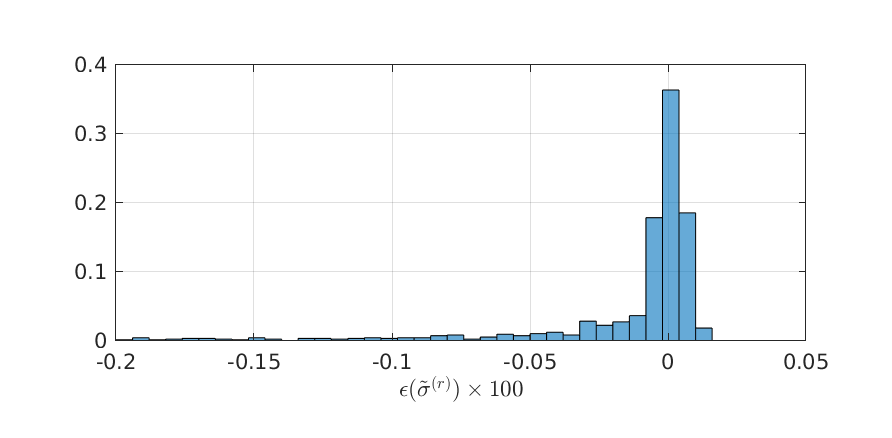}
\includegraphics[width=\columnwidth]{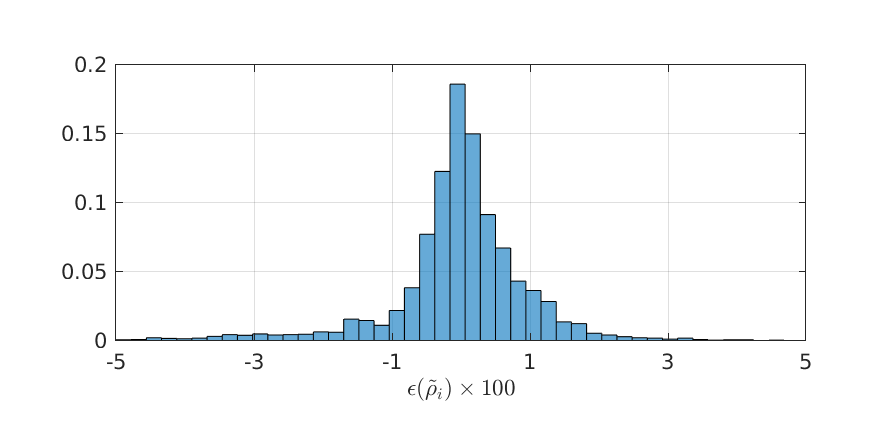}
\caption{Histograms of the relative deviation (in percentage) of the subscription ratios and fractions of subscribers with the proposed solution ($4$~NSTs, $20$~cells, $\alpha=3$, $\gamma^{(r)}\in[0.25,4]$).}\label{fig:hists2r20a5}
\end{center}
\end{figure}

These results suggest that the proposed solution provides a very good approximation of the equilibrium in a general case.
In order to ensure the validity of this approach for a wide range of the system parameters,
we have evaluated the relative deviations of the subscription ratio, $\epsilon(\tilde{\sigma}^{(r)})$, 
and of the fraction of subscribers, $\epsilon(\tilde{\rho}_i)$,
for a diversity of scenarios as described in Section~\ref{sec:setup}.

Fig.~\ref{fig:hists2r20a5} shows the histograms for the relative deviation (expressed as percentages) of 
the subscription ratio ($\epsilon(\tilde{\sigma}^{(r)})$) 
and of the fraction of subscribers ($\epsilon(\tilde{\rho}_i)$)
in one type of scenario
with $4$~NSTs, $20$~cells, $\alpha=3$ and $\gamma^{(r)}\in[0.25,4]$.
We see that the values of the relative deviation of the fraction of subscribers 
are distributed almost symmetrically around~0  and most values are below $1\,\%$.
The relative deviation of the subscription ratio is even smaller (most values are below $0.1\,\%$)
and that negative values are more frequent, which suggests that the proposed solution tends to provide an underestimate.

Finally, Table~\ref{table:KPI3e4} shows the 90th and 95th percentiles of the module of 
$\epsilon(\tilde{\rho}_i)$ and $\epsilon(\tilde{\sigma}^{(r)})$  for a wide range of scenarios.
It is observed that the deviations are slightly higher when $\gamma^{(r)}$  values are smaller (in the range $[0.25,0.5]$). 
But even in the worst case, 
the 95th percentile of $\vert\epsilon(\tilde{\rho}_i)\vert$ is below $4.2\,\%$ 
and the 95th percentile of $\vert\epsilon(\tilde{\sigma}^{(r)})\vert$ is below $0.32\,\%$.

\begin{table}
\centering
\caption{Percentiles of the module of $\epsilon(\tilde{\rho}_i)$ and $\epsilon(\tilde{\sigma}^{(r)})$  (in percentage)}
\resizebox{\columnwidth}{!}{
\begin{tabular}{rrrrrrrrr}
\toprule 
\multicolumn{8}{c}{\vspace{-4mm}} \\
\multicolumn{3}{c}{} &  &
 \multicolumn{2}{c}{$\vert\epsilon(\tilde{\rho}_i)\vert\times100$}  &&
\multicolumn{2}{c}{$\vert\epsilon(\tilde{\sigma}^{(r)})\vert\times100$} \\ 
 \cmidrule{5-6} \cmidrule{8-9}
$S$ & $R$ & $\alpha$ & $[\gamma_{\min},\gamma_{\max}$] &  $P_{90}$ & $P_{95}$ && $P_{90}$ & $P_{95}$ \\ \hline
 \multirow{4}{*}{$2$}& \multirow{4}{*}{$10$}  & $1$ &  \multirow{4}{*}{$[0.25,4]$}  & $1.5$ & $2.0$ && $0.112$ & $0.193$\\ 
  &  & $3$ &  & $2.7$ & $3.8$ && $0.112$ & $0.240$\\ 
  &  & $5$ &  & $2.6$ & $4.0$ && $0.093$ & $0.208$\\ 
  &  & $7$ &  & $2.1$ & $4.0$ && $0.051$ & $0.151$\\ \cmidrule{1-4}    
 \multirow{4}{*}{$2$} &  \multirow{4}{*}{$20$} & $1$ &  \multirow{4}{*}{$[0.25,4]$}   & $1.9$ & $2.7$ && $0.141$ & $0.311$\\ 
  &  & $3$ &  & $2.9$ & $4.1$ && $0.138$ & $0.284$\\ 
  &  & $5$ &  & $2.5$ & $3.8$ && $0.071$ & $0.174$\\ 
  &  & $7$ &  & $2.0$ & $2.9$ && $0.042$ & $0.106$\\ \cmidrule{1-4}      
 \multirow{4}{*}{$4$} &  \multirow{4}{*}{$20$}  & $1$ &  \multirow{4}{*}{$[0.25,4]$}   & $0.6$ & $0.8$ && $0.026$ & $0.050$\\ 
  &  & $3$ &  & $1.5$ & $2.2$ && $0.061$ & $0.131$\\ 
  &  & $5$ &  & $1.2$ & $1.8$ && $0.033$ & $0.064$\\ 
  &  & $7$ &  & $1.3$ & $2.1$ && $0.036$ & $0.071$\\   \cmidrule{1-4}   
 \multirow{4}{*}{$2$} & \multirow{4}{*}{$10$}  & \multirow{4}{*}{$3$}& $[0.25,0.5]$ & $0.4$ & $0.6$ && $0.039$ & $0.053$\\ 
  &  & & $[0,1]$ & $0.7$ & $1.2$ && $0.064$ & $0.082$\\ 
  &  & & $[1,2]$ & $1.0$ & $1.5$ && $0.038$ & $0.054$\\ 
  &  & & $[2,4]$ & $0.6$ & $0.9$ && $0.007$ & $0.011$\\   \cmidrule{1-4}    
 \multirow{4}{*}{$2$} &  \multirow{4}{*}{$20$}  & \multirow{4}{*}{$3$}& $[0.25,0.5]$ & $0.4$ & $0.6$ && $0.039$ & $0.055$\\ 
  &  & & $[0,1]$ & $1.1$ & $1.6$ && $0.082$ & $0.104$\\ 
  &  & & $[1,2]$ & $1.1$ & $1.9$ && $0.050$ & $0.075$\\ 
  &  & & $[2,4]$ & $0.6$ & $0.9$ && $0.007$ & $0.012$\\ \cmidrule{1-4}    
 \multirow{4}{*}{$4$} &  \multirow{4}{*}{$20$}  & \multirow{4}{*}{$3$}  & $[0.25,0.5]$  & $0.4$ & $0.5$ && $0.028$ & $0.036$\\ 
  &  & & $[0,1]$ & $0.7$ & $0.9$ && $0.038$ & $0.051$\\ 
  &  & & $[1,2]$ & $0.7$ & $0.9$ && $0.015$ & $0.023$\\ 
  &  & & $[2,4]$ & $0.3$ & $0.3$ && $0.005$ & $0.005$\\   
\bottomrule
\end{tabular}}\label{table:KPI3e4}
\end{table}

\section{Conclusions}\label{sec:conclusion}

In this work, a business model is proposed and analyzed where the NSTs provide mobile communications services to final users and provision themselves from a resource pool operated by an InP by means of network slicing mechanisms.
We propose a solution for the Nash equilibrium of the competition between the NSTs,  where each NST strategically distributes it share of the resources among its subscribers.
The proposed solution has the following properties:
\begin{itemize} 
\item It provides the exact values at the equilibrium if the cells are homogeneous in terms of normalized capacity. In this case, the subscription ratio is the same in all cells.
\item Otherwise, if the cells are heterogeneous, it provides an accurate approximation of the equilibrium.
\item In each cell, the subscription ratio increases with the normalized capacity, and
depends on the share distribution, being maximum when all shares are equal.
\item Each NST obtains a fraction of subscribers which is the same in all cells. It depends only on its share, the user sensitivity and the share distribution, and is minimum when the shares of all the other NSTs are equal.
\end{itemize}


\appendices


\bigskip
 

\section{Proofs}\label{sec:proofs}
\subsection{Lemmas}\label{sec:lemmas}

Let 
\begin{equation}\label{eq:f}
f_{a,\beta}(x)=x-a(1-x)^{1-\beta}, \qquad a\geq 0, \ 0\leq \beta < 1.
\end{equation}

\begin{lemma}\label{lemma:NLeq}
The function $f_{a,\beta}(x)$ has one, and only one, root in $[0,1)$.
Moreover, if $a>0$ the root lies in $(0,1)$.
\end{lemma}
\begin{IEEEproof}
If $a=0$ the proof is immediate, so we focus on the case $a>0$.

Clearly, $f_{a,\beta}(x)$ is continuous in $[0,1]$, 
$f_{a,\beta}(0)=-a<0$ and $f_{a,\beta}(1)=1>0$.
Thus, $f_{a,\beta}(x)$ has a root in $(0,1)$.
Furthermore, 
\begin{equation}
  f'_{a,\beta}(x)=1+\frac{a(1-\beta)}{(1-x)^\beta} > 0, \qquad x\in (0,1),
\end{equation}
in other words, $f_{a,\beta}(x)$ is increasing in $(0,1)$.
Consequently, $f_{a,\beta}(x)$ has a unique root in $(0,1)$.
\end{IEEEproof}

Let $z(a,\beta)$ be the root of $f_{a,\beta}(x)$ in $[0,1)$.
Lemma~\ref{lemma:NLeq} guarantees that the function $z(a,\beta)$ is well-defined.
Moreover, since 
\begin{equation}
 \frac{\partial f_{a,\beta}(x)}{\partial x} = 1+\frac{a(1-\beta)}{\left(1-x\right)^\beta}  > 0,
\end{equation}
for all $x \in [0,1)$, $a\geq 0$ and $0\leq \beta < 1$,
by the implicit function theorem it follows that 
$z(a,\beta)$ is a continuously differentiable function.
\begin{lemma}\label{lemma:inc_a}
The function $z(a,\beta)$ is increasing with respect to $a$.
\end{lemma}
\begin{IEEEproof}
From $z-a(1-z)^{1-\beta}=0$, 
it follows immediately that
\begin{equation}
 \frac{\partial z}{\partial a} = \left(1+\frac{a(1-\beta)}{\left(1-z\right)^\beta}\right)^{-1}  > 0.
\end{equation}
\end{IEEEproof}

\begin{lemma}\label{lemma:max_sum_xi_beta}
Let 
\begin{equation}
	g(\vec{x})=x_1^\beta+\cdots+x_n^\beta, 
\end{equation}
with $\beta<1$, 
defined in the domain 
\begin{equation}
	D=\{(x_1,\ldots, x_n) : x_i\geq0; \ x_1+\cdots+x_n=1\}.
\end{equation}

Then,
\begin{equation}
	\max_D g(\vec{x})=g(\vec{x}^*)=n^{1-\beta},
\end{equation}
where $\vec{x}^* = (1/n,\ldots,1/n)$ is the unique maximum point in $D$,
that is, 
$g(\vec{x})< g(\vec{x}^*)$ for all $\vec{x} \in D \smallsetminus \{\vec{x}^*\}$.
\end{lemma}
\begin{IEEEproof}
We first consider the maximization in $D_0=\{(x_1,\ldots, x_n) : x_i>0; \ x_1+\cdots+x_n=1\} \subset D$.
It is easy to check that $g(\vec{x})$ is strictly concave in $D_0$.
Thus, the KKT conditions are both necessary and sufficient for a point $\vec{x}$ to be a maximum in $D_0$.

The KKT conditions are 
\begin{align}
 \frac{\partial g}{\partial x_i} = \frac{\beta}{x_i^{1-\beta}} &= \lambda, \\
  x_1+\cdots+x_n &= 1,
\end{align}
whose only solution is $x_1=\cdots=x_n=1/n$,
and hence $\max_{D_0} g(\vec{x})=n^{1-\beta}$.

Now suppose that $k$ of the $x_i$'s are zero, with $1\leq k \leq n$. 
Assume, without any loss of generality, they are the first $k$, and let
$D_k=\{(x_1,\ldots, x_n) : x_1=\cdots=x_k=0;\ x_i>0, i=k+1,\ldots,n; \ x_1+\cdots+x_n=1\} 
\subset D \smallsetminus D_0$

Applying the above result for the maximization over $D_0$ to the maximization over $D_k$,
we may conclude that 
$\max_{D_k} g(\vec{x})=(n-k)^{1-\beta} < n^{1-\beta} = \max_{D_0} g(\vec{x})$.

Therefore, we can assert that 
$\max_D g(\vec{x})= \max_{D_0} g(\vec{x})=g(\vec{x}^*)=n^{1-\beta}$.
\end{IEEEproof}

\begin{lemma}\label{lemma:min_sum_xi_beta}
Let 
\begin{equation}
	g(\vec{x})=x_1^\beta+\cdots+x_n^\beta, 
\end{equation}
with $\beta<1$, 
defined in the domain 
\begin{equation}
	D=\{(x_1,\ldots, x_n) : x_i\geq0; \ x_1+\cdots+x_n=1\}.
\end{equation}

Then,
\begin{equation}
	\min_D g(\vec{x})=g(\vec{x}^*)=1.
\end{equation}
\end{lemma}
\begin{IEEEproof}
Since $\beta<1$, it is clear that $g(\vec{x}) \geq  x_1+\cdots+x_n=1$.
Furthermore, $g(1,0,\ldots,0)= \cdots = g(0,0,\ldots,1) =1$.
Therefore,  $\min_D g(\vec{x})=1$.
\end{IEEEproof}

\subsection{Proof of Proposition~\ref{prop:penetration}}

When $r_0^{(j)}>0$, we can rewrite~\eqref{eq:usersFunWeights1} as
\begin{multline}\label{eq:usersFunWeights4}
\left(\frac{\omega^{(j)}_i}{n^{(j)}_i}\right)^{\alpha}
= \frac{n^{(j)}_i}{n^{(j)}} 
\left(
  \sum_{t \in \mathcal{S}}\left(\frac{\omega^{(j)}_t}{n^{(j)}_t}\right)^{\alpha}
   + 
  \left(\frac{\sum_{t \in \mathcal{S}} \omega^{(j)}_t}{n^{(j)} \ncap{j}}  \right)^{\alpha}
\right),
\\
i \in \mathcal{S}, \quad j \in \mathcal{B}.
\end{multline}
Now, adding~\eqref{eq:usersFunWeights4} over all $i \in \mathcal{S}$ yields
\begin{multline}\label{eq:sumUsersFunWeights}
\sum_{t \in \mathcal{S}}\left(\frac{\omega^{(j)}_t}{n^{(j)}_t}\right)^{\alpha}
= 
\penratio{j} 
  \sum_{t \in \mathcal{S}}\left(\frac{\omega^{(j)}_t}{n^{(j)}_t}\right)^{\alpha}
   + 
\penratio{j} 	
  \left(\frac{\sum_{t \in \mathcal{S}} \omega^{(j)}_t}{n^{(j)} \ncap{j}} \right)^{\alpha},
\\
j \in \mathcal{B},
\end{multline}
which can be rewritten as
\begin{equation}\label{eq:sumUsersFunWeights2}
\sum_{t \in \mathcal{S}}\left(\frac{\omega^{(j)}_t}{n^{(j)}_t}\right)^{\alpha}
=
  \frac{\penratio{j}}{1-\penratio{j}}  
  \left(\frac{\sum_{t \in \mathcal{S}} \omega^{(j)}_t}{n^{(j)} \ncap{j}}  \right)^{\alpha},
\quad j \in \mathcal{B}.
\end{equation}

Substituting~\eqref{eq:sumUsersFunWeights2} into~\eqref{eq:usersFunWeights4} and solving for $n^{(j)}_i$ yields
\begin{multline}\label{eq:ns}
\frac{n^{(j)}_i}{n^{(j)}}
=
\left(1-\penratio{j}\right)^{1-\beta} 
\left(\ncap{j}\right)^\beta
\frac{\left(\omega^{(j)}_i\right)^{\beta}}
     {\left(\sum_{t \in \mathcal{S}} \omega^{(j)}_t\right)^{\beta}},
\\	
i \in \mathcal{S}, \quad j \in \mathcal{B},
\end{multline}
where $\beta = \alpha/(\alpha+1)<1$. 

Finally, adding~\eqref{eq:ns} over all $i \in \mathcal{S}$, we obtain
\begin{equation}\label{eq:penetration_NLeq_proof}
\penratio{j} - \left(\ncap{j}\right)^\beta
               \frac{\sum_{t \in \mathcal{S}} \left(\omega^{(j)}_t\right)^{\beta}}
                    {\left(\sum_{t \in \mathcal{S}} \omega^{(j)}_t\right)^{\beta}}
               \left(1-\penratio{j}\right)^{1-\beta} = 0,
  \qquad j \in \mathcal{B}.
\end{equation}
The application of Lemma~\ref{lemma:NLeq} guarantees that~\eqref{eq:penetration_NLeq_proof}
has a unique solution in $(0,1)$.

In the case $r_0^{(j)}=0$, we can rewrite~\eqref{eq:usersFunWeights1} as
\begin{equation}
\left(\frac{\omega^{(j)}_i}{n^{(j)}_i}\right)^{\alpha}
=
\frac{n^{(j)}_i}{n^{(j)}} 
  \sum_{t \in \mathcal{S}}\left(\frac{\omega^{(j)}_t}{n^{(j)}_t}\right)^{\alpha},
\qquad i \in \mathcal{S}, \quad j \in \mathcal{B}.
\end{equation}
Adding this equality over all $i \in \mathcal{S}$ immediately leads to $\penratio{j}=1$.

\subsection{Proof of Proposition~\ref{prop:penetration_normCapacity}}

The first part of the proposition follows immediately by applying Lemma~\ref{lemma:inc_a}
with 
\begin{equation}\label{eq:def_a}
	a=  \left(\ncap{j}\right)^\beta
    \frac{\sum_{t \in \mathcal{S}} \left(\omega^{(j)}_t\right)^{\beta}}
         {\left(\sum_{t \in \mathcal{S}} \omega^{(j)}_t\right)^{\beta}},
\end{equation}
which clearly increases with $\ncap{j}$.

From~\eqref{eq:def_a} it follows that
\begin{equation}
  \lim_{\gamma^{(j)}\rightarrow 0} \penratio{j}  
   = \lim_{a\rightarrow 0} z(a,\beta) = z(0,\beta) =0.
\end{equation}

Now, from the definition of $z(a,\beta)$  and~\eqref{eq:f} we have
\begin{equation}
  a = \frac{z(a,\beta)}{\big(1-z(a,\beta)\big)^{1-\beta}} 
	  < \frac{1}{1-z(a,\beta)}.		
\end{equation}

Thus, $1-1/a < z(a,\beta) < 1$, and consequently,
\begin{equation}
  \lim_{\gamma^{(j)}\rightarrow \infty} \penratio{j}  
   = \lim_{a\rightarrow \infty} z(a,\beta) = 1.
\end{equation}
\subsection{Proof of Proposition~\ref{prop:penetration_sensitivity}}
We first show that
\begin{align}
  \label{eq:lim_z_beta_0}
  \lim_{\beta\rightarrow 0} z(a,\beta) &= \frac{a}{1+a}, \\
  \label{eq:lim_z_beta_1}
  \lim_{\beta\rightarrow 1} z(a,\beta) &= \min(1,a). 
\end{align}

It is clear that
\begin{equation}
  \lim_{\beta\rightarrow 0} f_{a,\beta}\Big(\frac{a}{1+a}\Big)
   = \lim_{\beta\rightarrow 0} \bigg( \frac{a}{1+a} - \frac{a}{(1+a)^{1-\beta}}\bigg) = 0.
\end{equation}
Hence the result in~\eqref{eq:lim_z_beta_0} follows from the continuity of $f_{a,\beta}(x)$
and the definition of $z(a,\beta)$.

Likewise, if $a<1$, we have
\begin{equation}
  \lim_{\beta\rightarrow 1} f_{a,\beta}(a)
   = \lim_{\beta\rightarrow 1} \big( a - a(1-a)^{1-\beta}\big) = 0,
\end{equation}
and hence 
\begin{equation}\label{eq:lim_z_beta_1_a_less_1}
  \lim_{\beta\rightarrow 1} z(a,\beta) = a, \qquad \text{if $a<1$.} 
\end{equation}

We now turn to the case $a=1$.
For every $\varepsilon > 0$ (we assume without restriction of generality that $\varepsilon < 1$),
we take $\delta=\log(1-\varepsilon)/\log(\varepsilon)>0$ such that
if $|1-\beta|= 1-\beta < \delta$, we have
\begin{multline}
  f_{a,\beta}(1-\varepsilon) = 1-\varepsilon - \varepsilon^{1-\beta} \\
                             <  1-\varepsilon - \varepsilon^{\frac{\log(1-\varepsilon)}{\log(\varepsilon)}}
                             = 1-\varepsilon - (1-\varepsilon) = 0.
\end{multline}
We thus have $f_{a,\beta}(1-\varepsilon)<0$ and $f_{a,\beta}(1)=1>0$. 
Consequently, $1-\varepsilon < z(a,\beta)  < 1$, and therefore $|z(a,\beta) -1| < \varepsilon$,
which proves that
\begin{equation}\label{eq:lim_z_beta_1_a_equal_1}
  \lim_{\beta\rightarrow 1} z(1,\beta) = 1.
\end{equation}

Consider now the case $a>1$.
Since
$f\big(1-a^{-1/(1-\beta)}\big) = -a^{-1/(1-\beta)} < 0$,
we have 
\begin{equation}
1-a^{\frac{-1}{1-\beta}} < z(a,\beta) < 1.
\end{equation}
This, together with the fact that
\begin{equation}
\lim_{\beta\rightarrow 1}  1- a^{\frac{-1}{1-\beta}} = 1.
\end{equation}
yields
\begin{equation}\label{eq:lim_z_beta_1_a_greater_1}
  \lim_{\beta\rightarrow 1} z(a,\beta) = 1, \qquad \text{if $a>1$.} 
\end{equation}

Finally, 
combining~\eqref{eq:lim_z_beta_1_a_less_1},
\eqref{eq:lim_z_beta_1_a_equal_1}  
and~\eqref{eq:lim_z_beta_1_a_greater_1} 
we obtain the result in~\eqref{eq:lim_z_beta_1}.

Now we set $a$ as in~\eqref{eq:def_a} and note that
\begin{align}
  \label{eq:lim_a_beta_0}
  \lim_{\beta\rightarrow 0} a &= \cardS, \\
  \label{eq:lim_a_beta_1}
  \lim_{\beta\rightarrow 1} a &= \ncap{j}. 
\end{align}
These, together with~\eqref{eq:lim_z_beta_0} and~\eqref{eq:lim_z_beta_1},
and the continuity of $z(a,\beta)$ 
completes the proof of the proposition.
\subsection{Proof of Proposition~\ref{prop:n_rs}}

The expression~\eqref{eq:n_rs} 
follows immediately by combining~\eqref{eq:rho_ij} with~\eqref{eq:ns} and~\eqref{eq:penetration_NLeq_proof}.

\subsection{Proof of Proposition~\ref{prop:penetration_minMax}}

The proof is straightforward by applying Lemma~\ref{lemma:inc_a},
with 
\begin{equation}
	a=  \left(\ncap{j}\right)^\beta \sum_{t \in \mathcal{S}} s_t^{\beta},
\end{equation}
along with Lemma~\ref{lemma:max_sum_xi_beta} and Lemma~\ref{lemma:min_sum_xi_beta}.
\subsection{Proof of Proposition~\ref{prop:equalPenetrations}}

The first conclusion was justified in the discussion preceding the proposition,
and the second is obtained by substituting~\eqref{eq:penetration_ref} into~\eqref{eq:genSolWeights}.

\subsection{Proof of Proposition~\ref{prop:derivatives}}
Taking derivatives of~\eqref{eq:profitsweightsgeneral} yields
\begin{align}
     \label{eq:Dj_proof}
	 \frac{\partial \Pi_i}{\partial \omega^{(j)}_i} &= 
	     n^{(j)} \frac{\partial \penratio{j}}{\partial \omega^{(j)}_i} \rho_i^{(j)}
		 +
	     n^{(j)} \penratio{j} \frac{\partial \rho_i^{(j)}}{\partial \omega^{(j)}_i} 	
	 \\ \nonumber
	 \frac{\partial^2 \Pi_i}{\partial \big(\omega^{(j)}_i\big)^2} &=
	     n^{(j)} \frac{\partial^2 \penratio{j}}{\partial \big(\omega^{(j)}_i\big)^2} \rho_i^{(j)}
		 \\ \label{eq:Djj_proof}
		 & \quad +
	     2n^{(j)} \frac{\partial \penratio{j}}{\partial\omega^{(j)}_i} 
		          \frac{\partial\rho_i^{(j)}}{\partial\omega^{(j)}_i} 
		 +
	     n^{(j)}\penratio{j} \frac{\partial^2 \rho_i^{(j)}}{\partial \big(\omega^{(j)}_i\big)^2} 		
\end{align}

The derivatives of $\rho_i^{(j)}$ and $\penratio{j}$ are obtained 
from~\eqref{eq:n_rs} and~\eqref{eq:penetration_NLeq}, respectively.
Note that in the latter case implicit differentiation must be used.

Clearly, $\rho_i^{(j)}$ and $\penratio{j}$ are independent of
$\omega^{(k)}_i$ for all $k\in \mathcal{B} \setminus \{j\}, \ i \in \mathcal{S}$
and, thus,
\begin{equation}
		 \frac{\partial^2 \Pi_i}{\partial\omega^{(j)}_i \partial\omega^{(k)}_i} = 0.
\end{equation}

After some algebra we get
\begin{align}
	 \frac{\partial \rho_i^{(j)}}{\partial \omega^{(j)}_i} &= 	
	     \frac{\beta}{\omega^{(j)}_i} \rho_i^{(j)} (1-\rho_i^{(j)}),
	 \\	
	 \frac{\partial \penratio{j}}{\partial \omega^{(j)}_i} &= 	
	     \frac{1-\penratio{j}}{1-\beta\penratio{j}} \frac{\beta\penratio{j}}{\omega^{(j)}_i} 
		 (\rho_i^{(j)}-x_i^{(j)}),
	 \\	
	 \frac{\partial^2 \rho_i^{(j)}}{\partial \big(\omega^{(j)}_i\big)^2} &=
	     \frac{-\beta}{\big(\omega^{(j)}_i\big)^2} \rho_i^{(j)} (1-\rho_i^{(j)}) (1-\beta+2\beta\rho_i^{(j)})
\end{align}
\begin{multline}
	 \frac{\partial^2 \penratio{j}}{\partial \big(\omega^{(j)}_i\big)^2} =			
	     \frac{1-\penratio{j}}{1-\beta\penratio{j}} \frac{\beta\penratio{j}}{\big(\omega^{(j)}_i\big)^2} 
		   \Bigg(\frac{1-2\penratio{j}+\beta\big(\penratio{j}\big)^2}{\big(1-\beta\penratio{j}\big)^2}
			\\	
			\quad \times \beta (\rho_i^{(j)}-x_i^{(j)}) 
				+  \beta \rho_i^{(j)} (1-\rho_i^{(j)}) - \rho_i^{(j)} + \big(x_i^{(j)}\big)^2
		   \Bigg)	
\end{multline}
Finally, substituting these expressions into~\eqref{eq:Dj_proof} and~\eqref{eq:Djj_proof}
we obtain, after some algebra, \eqref{eq:Dj} and~\eqref{eq:Djj}.
	
\subsection{Proof of Proposition~\ref{prop:psKKTnecCond}}
From the expression of the weights $\omega^{(j)}_i$, see~\eqref{eq:solWeights_ref}, 
it is clear that $\sum_{j \in \mathcal{B}} \omega^{(j)}_i = s_i$, 
and thus the conditions~\eqref{eq:feasibility} and~\eqref{eq:complementarySlackness} are satisfied.
	
Since the condition of Proposition~\ref{PROP:EQUALPENETRATIONS} is met, 
we have $\penratio{j} = \sigma$,
and from the expression of the weights it follows that
$\rho_i^{(j)}= \rho_i$, where $\rho_i$ is given in~\eqref{eq:rho_i},
and $x_i^{(j)} = s_i$.
Substituting these into~\eqref{eq:Dj} 
and recalling that~\eqref{eq:stationary} is equivalent to~\eqref{eq:br2}, 
we obtain~\eqref{eq:multiplier}.
Finally, from~\eqref{eq:multiplier} it is clear that $\mu_i>0$ 
and therefore~\eqref{eq:posMultiplier} is satisfied.



\bibliographystyle{IEEEtran}



\begin{IEEEbiography}[{\includegraphics[width=1in,height=1.25in,clip,keepaspectratio]{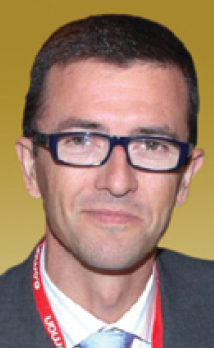}}]{Luis Guijarro} received the M.Eng. and Ph.D. degrees in telecommunications from the Universitat Polit\`ecnica de Val\`encia (UPV), Valencia, Spain. He is currently a Professor in telecommunications economics and regulation with the Department of Communications, UPV. He has coauthored the book Electronic Communications Policy of the European Union (UPV, 2010). He has researched in traffic management in ATM networks and in e-Government. His current research area is economic modeling of telecommunication service provision. He has contributed in the topics of peer-to-peer interconnection, cognitive radio networks, net neutrality, wireless sensor networks, and 5G. 
\end{IEEEbiography}

\begin{IEEEbiography}[{\includegraphics[width=1in,height=1.25in,clip,keepaspectratio]{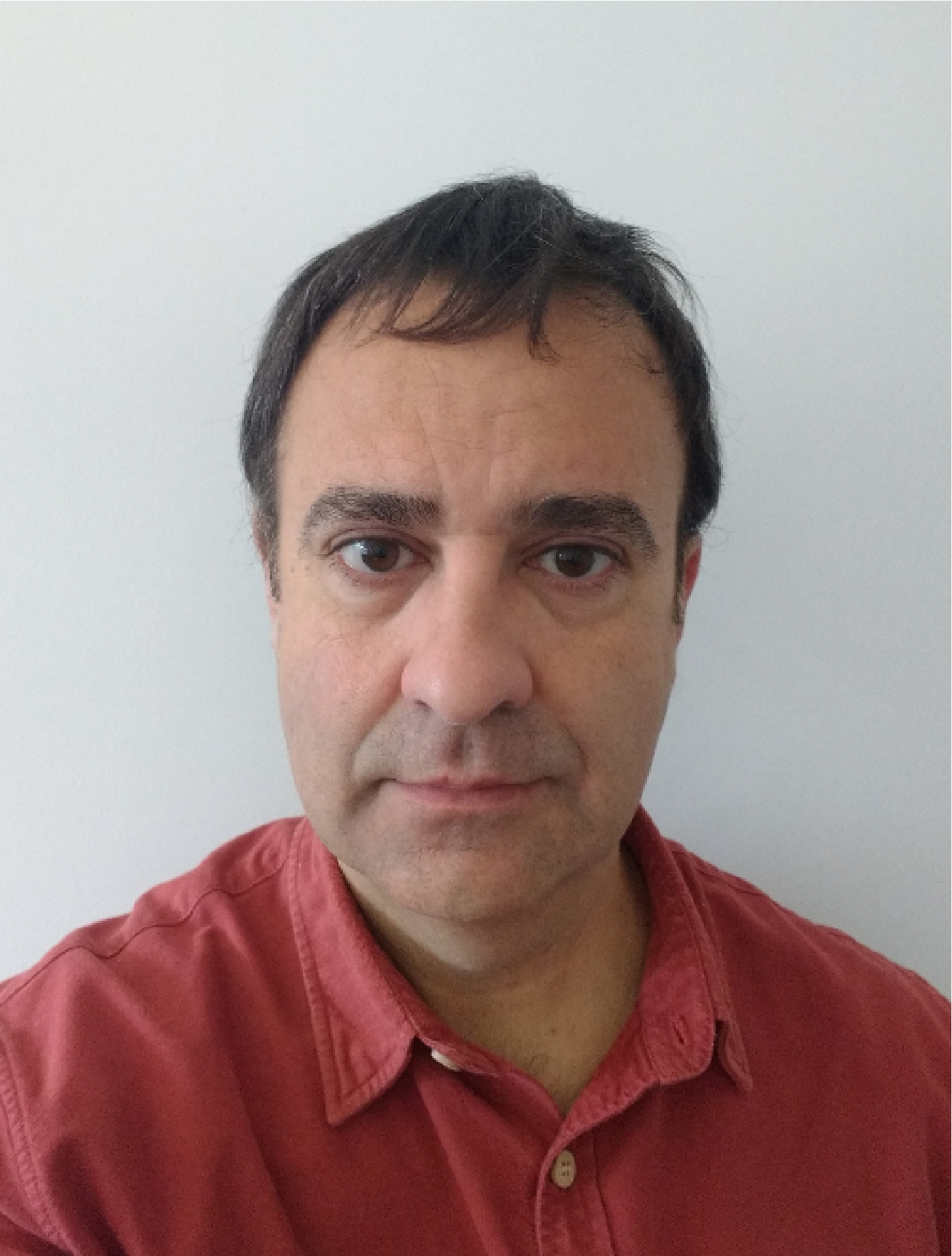}}]{Jos\'e-Ram\'on Vidal} received the Ph.D. degree in telecommunication engineering from the Universitat Polit\`ecnica de Val\`encia (UPV), Valencia, Spain. He is currently an Associate Professor with the Department of Communications, Higher Technical School of Telecommunication Engineering, UPV. His current research interest includes the application of game theory to resource allocation in cognitive radio networks and to economic modeling of telecommunication service provision. In these areas, he has authored or coauthored several papers in refereed journals and conference proceedings
\end{IEEEbiography}

\begin{IEEEbiography}[{\includegraphics[width=1in,height=1.25in,clip,keepaspectratio]{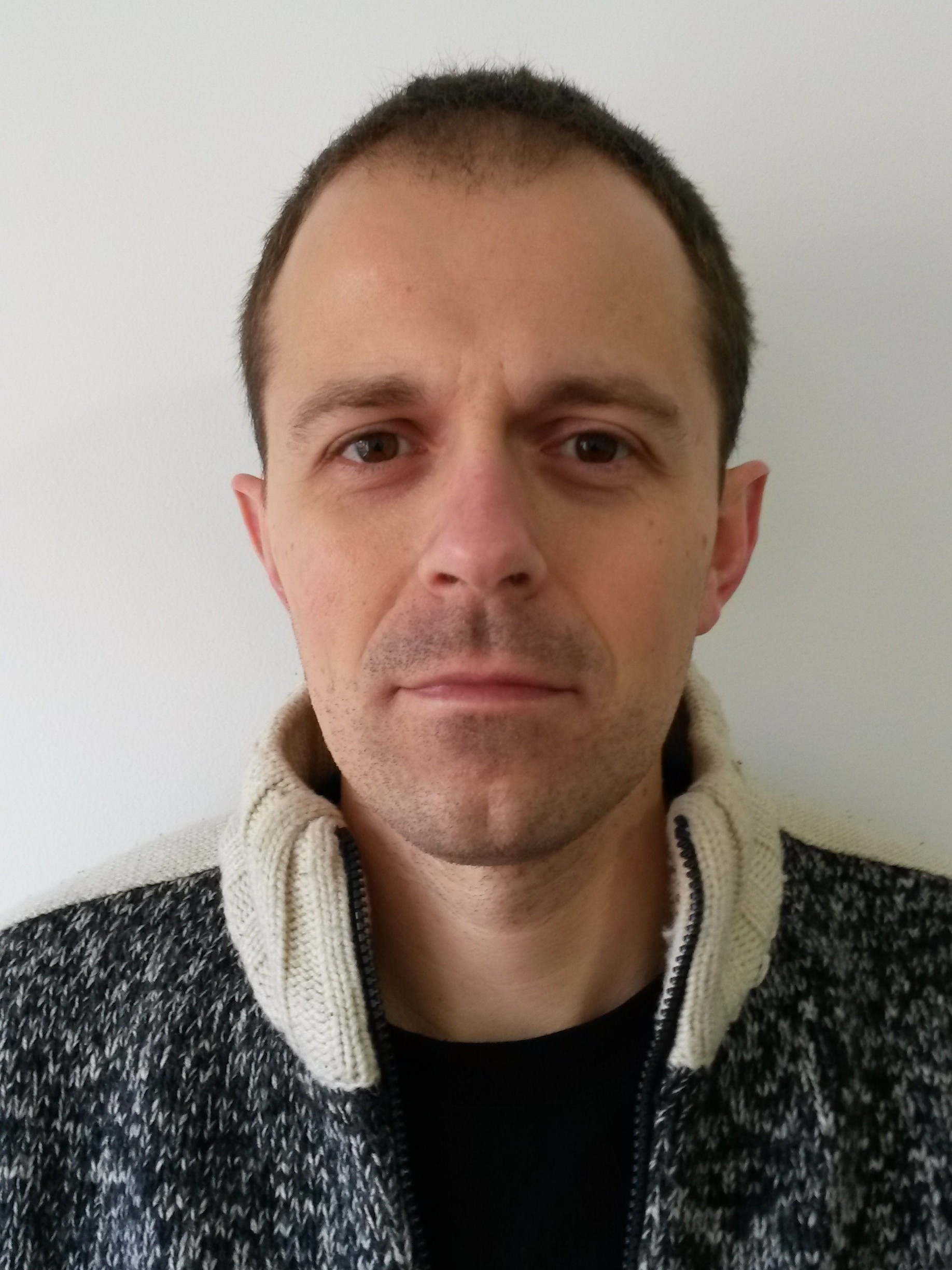}}]{Vicent Pla}
received the B.E. and M.E. degrees in telecommunication engineering and the Ph.D. degree from the Universitat Polit\`ecnica de Val\`encia (UPV), Valencia, Spain, in 1997 and 2005, respectively, and the B.Sc. degree in mathematics from the Universidad Nacional de Educaci\'on a Distancia, Madrid, Spain, in 2015. In 1999, he joined the Department of Communications, UPV, where he is currently a Professor. His research interest includes modeling and performance analysis of communication networks. During the past few years, most of his research activity has focused on traffic and resource management in wireless networks. In these years, he has authored or coauthored numerous papers in refereed journals and conference proceedings and has been an active participant in several research projects.
\end{IEEEbiography}

\EOD

\end{document}